\documentclass[useAMS,usenatbib]{mn2e}

\usepackage{epsfig}
\usepackage{graphicx}
\usepackage{amsmath}

\title[Searching for evidence of radiative line driving]{Searching for the signature of radiative line driving : On the absence of
Ly~$\alpha$--N~{\sc v} line-locking features in a large sample of BALQSOs}
\author[C. E. Cottis$^{1}$, M. R. Goad$^{1}$,
C. Knigge$^{2}$, and S. Scaringi$^{3}$]{C. E. Cottis\thanks{E-mail:
cec17@star.le.ac.uk}$^{1}$, M. R. Goad$^{1}$, C. Knigge$^{2}$, and
S. Scaringi$^{3}$\\ $^{1}$ Department of Physics and Astronomy, University of
Leicester, University Road, LE1 7RH, UK.\\ $^{2}$ School of Physics and
Astronomy, University of Southampton, Southampton SO18 1BT, UK. \\ $^{3}$ 
Department of Astrophysics, IMAPP, Radboud University Nijmegen, P.O. Box
9010, 6500 GL Nijmegen, The Netherlands.}
\begin{document}

\date{}

\pagerange{\pageref{firstpage}--\pageref{lastpage}} \pubyear{2010}

\maketitle

\label{firstpage}

\begin{abstract}

We have searched the hybrid BALQSO catalogue of Scaringi et~al.  derived from
data release 5 of the Sloan Digital Sky Survey in order to compile the largest
sample of objects displaying spectral signatures which may be indicative of
radiative line driving.  The feature in question is the ``ghost of
Ly-$\alpha$'', a line-locking feature previously identified in the broad
C~{\sc iv} and Si~{\sc iv} absorption lines of a small fraction of BALQSOs,
and formed via the interaction of Ly-$\alpha$ photons with N~{\sc v}
ions.

We test, where possible the criteria required to produce an observable ghost
feature. These criteria include: significant broad absorption, strong
intrinsic Ly-$\alpha$ emission, narrow Ly-$\alpha$, strong N~{\sc v}
absorption, and a weak far-UV continuum. No single ghost-candidate meets all
of these criteria. Furthermore, we find that these criteria are not met
significantly more often in ghost-candidates than in a comparison sample
chosen to exhibit relatively featureless broad absorption troughs. Indeed, the
only significant differences we find between our ghost-candidate and
comparison samples, is that on average, our ghost-candidate sample displays
(i) significantly stronger N~{\sc v} absorption, and (ii) the onset of
absorption occurs at lower velocities in our ghost-candidate objects.

Significantly, we find no evidence for an excess of objects whose absorption
troughs bracket the location of the Ly-$\alpha$--N{\sc v} line-locking region,
rather the location of ghost-like features appears to be independent of any
systematic velocity, with comparable numbers appearing both redward and
blueward of the ghost-zone.  Thus, the majority of objects identified here as
strong ghost-candidates are likely multi-trough interlopers whose absorption
feature simply bracket the region of interest.

\end{abstract}

\section{Introduction}

Broad absorption line quasars (BALQSOs) as their name suggests, show strong
broad blue-shifted absorption lines in their spectra believed to be indicative
of high velocity ($\sim$0.1c) out-flowing winds. BALQSOs represent
approximately 15\% of quasars in general
\citep{Reich03a,Trump06,Knigge,Sca09}. The majority ($\sim$ 85\%)
\citep{Spray,Reich03b} of BALQSOs are known as HiBALs, displaying absorption
in lines of high ionisation only (e.g. N~{\sc v}, Si~{\sc iv}, and C~{\sc
iv}). The remainder are classified as LoBALs and show in addition broad
absorption in lines of low ionisation species, most notably Al~{\sc iii} and
Mg~{\sc ii}. LoBALs are further sub-classified according to the presence of Fe
absorption, the Fe~LoBals. Since the spectra of LoBals are in general redder
than HiBals, \cite{Becker00} suggests that BALQSOs and in particular Fe~LoBALs
may be an early phase in the development of emerging or re-fuelled quasars.

There has been much debate about the relationship between BALQSOs and the
general quasar population as a whole. Simple unification schemes, suggest that
BALQSOs and non-BALQSOs are similar objects and that any observed differences
in their spectra arise due to orientation effects
\citep{Ogle99,Wey91,Sch99,Elvis}. In this picture, the relative fraction of
BALQSOs to non-BALQSOs has a simple geometric interpretation, representing the
fraction of sky, as seen from the source, obscured by gas (ie. the source
covering fraction), which in the simplest case, can be related to the
flow geometry (e.g. opening angle).

In recent years interest in BALQSOs outflows has risen sharply, principally
because of the realisation that such high velocity outflows carry a
substantial amount of energy and momentum into the ISM, and may therefore be
important in driving AGN feedback as well as providing a mechanism for
quenching star formation \citep{Scan}.  Indeed, the discovery of highly
ionised, very high-velocity X-ray outflows (e.g. Pounds et~al. 2003a, 2003b;
Pounds and Reeves 2009), for which the energy transport (in terms of
mechanical energy) is large enough to interrupt the growth of the host galaxy,
may provide the causal link behind the well-known correlation between the mass
of the central black hole and the mass of the bulge (e.g. Ferrarese and
Merritt 2000; Gebhardt et~al.  2000; Tremaine et~al. 2002).

However, despite the increasing importance of these outflows the precise
mechanism responsible for accelerating them to high velocity remains
uncertain, with radiative acceleration
\citep{Shlos85,AravI,Murr95,Prog00,Chel01} , Magneto Hydro Dynamic (MHD)
driven winds \citep{blandford,Em92,Kon94,Bot97} and thermally driven winds
seen as the main contenders (see e.g. \cite{Prog07} for a review). A major
hindrance to progress in this area is the absence of a clean discriminatory
observational signature of what are essentially orthogonal wind geometries.
For radio-quiet objects, AGN unification schemes tend to favour equatorial
wind geometries (e.g. Elvis 2000). By contrast, observations of radio-loud
BALQSOs \citep{Becker97,Becker00,brotherton98} and in particular polarisation
observations of PKS~0040-005 indicate a non-equatorial BAL outflow
\citep{brotherton06}. One possible explanation for the apparent difference in
wind geometry between radio-loud and radio-quiet objects, is that different
acceleration mechanisms operate in these two classes of objects.

Perhaps the strongest indicator that radiative driving is responsible for
accelerating at least some of the high velocity flows, is the appearance of
line-locking features in the spectra of a small fraction of BAL quasars
(e.g. Turnshek et~al. 1988, Weymann et~al. 1991, Korista et~al. 1993, Arav
1996, Vilkoviskij and Irwin 2001). The most well-studied of these is the
so-called ghost of Ly-$\alpha$, a small hump seen in the absorption troughs of
a small fraction (less than a few percent) of BALQSOs formed via the
interaction between Ly-$\alpha$ photons and N~{\sc v} ions (see e.g. Korista
et~al. 1993, Arav et~al.  1995, Arav 1996, and references therein).  However,
the reality of ghosts remains very much an open question. Previous studies
have been limited to small samples of relatively low quality spectra, with
strong ghosts often only becoming apparent in composite spectra (Arav
et~al. 1995, North et~al. 2006). Moreover, distinguishing between line-locking
features and similar features caused by the chance alignment of multiple
absorption systems is difficult with relatively small samples. Indeed, Korista
et~al. 1993, showed that in a sample 72 objects, evidence for line-locking
features was merely suggestive rather than convincing.

The aim of this paper is twofold: (i) to compile the largest sample of
uniformly selected ghost-candidate spectra, and (ii) attempt to
determine the origin of their ghost features, by testing on a source
by source basis, whether they meet the original criteria as set out by
Arav (1995) necessary for ghost-features to be observed.  In \S2
we describe the mechanism proposed for ghost formation and outline
Arav's criteria for the production of strong ghost signatures.
Selection of those objects comprising our ghost-candidate spectra and
our non-ghost control sample is described in \S3.  In \S4 we present
the results of testing each of the spectra in our ghost-candidate and
comparison samples against each of the criteria in turn necessary for
the formation of an observable ghost feature. We discuss the
implications of our findings in \S5.  Our conclusions are summarised
in \S6.

\section{The ghost of Ly-$\alpha$}

The feature that is poetically known as the ghost of Ly-$\alpha$ is a hump
seen in the absorption trough of a small fraction (less than 5\%, Arav 1995,
North et~al. 2006) of BALQSOs located 5900~km~s$^{-1}$ blue-ward of the centre
of the line to which the absorption is attributed. The feature was first
investigated by \cite{Kor93} after being seen in a difference spectrum of
BALQSO and non-BALQSO spectral composites produced by \cite{Wey91} who pointed
out that the difference between the two absorption troughs was approximately
equal to the velocity separation of N~{\sc v} and Ly-$\alpha$. In a series of
papers in the mid 90's \citep{AravI,AravII,AravIII,AravNature,AravV} Arav
suggested that the feature seen at this velocity was a result of the increased
radiative pressure on N~{\sc v} ions at this velocity within the outflow. In
this picture N~{\sc v} ions moving at 5900~km~s$^{-1}$ 'see' the Ly-$\alpha$
emission from the AGN at the energy of their own emission resulting in a large
increase in the scattering cross section of the Ly-$\alpha$
photons. Consequently, out-flowing N~{\sc v} ions receive an increase in
radiation pressure leading to a large injection of momentum into the
outflow. The out-flowing ions of other species are dragged to higher
velocities by electromagnetic interactions with the N~{\sc v} ions, leading to
a deficit of material at this velocity (because each out-flowing ion spends
very little time at this velocity before being accelerated to greater
velocities). In turn this causes a decrease in the opacity at this velocity
which manifests as a peak within the absorption troughs of the out-flowing
species.

This simple mechanism explains both the presence of the feature and why it's
position is linked to the rest frame of the source.\footnote{This is not the
case for shadowing of N~{\sc v} by Ly-$\alpha$ \citep{Kor93} where the
separation of the troughs is constant but their velocities are determined by
the velocity of the first trough.} \cite{AravV} suggested that the reason for
the feature being most prominent in C~{\sc iv} absorption rather than in other
species, is that the doublet separation of C~{\sc iv} is far smaller
(498~km~s$^{-1}$). By comparison the larger doublet separations of Si~{\sc iv}
(1933~km~s$^{-1}$) and O~{\sc vi} (1647~km~s$^{-1}$) result in the feature
being blurred over several thousand km~s$^{-1}$, producing a low contrast
feature which is difficult to detect in low S/N data. For this reason, in this
study we confine our search for ghost features to the C~{\sc iv} broad
absorption line only.

% why is this signature not ubiquitous
Since Ly-$\alpha$ and N~{\sc v} are amongst the strongest UV lines in AGN
spectra, an obvious question to ask is why aren't ghost-features ubiquitous
amongst the general BALQSO population? To answer this, Arav proposed a set of
physically motivated criteria which had to be met before ghost features would
be observable. These are:

\begin{enumerate}
\item Significant broad absorption in the region between 3000 and
9000~km~s$^{-1}$ blue-ward of line centre,
\item Strong intrinsic Ly-$\alpha$ emission,
\item Narrow Ly-$\alpha$ emission,
\item Strong broad N~{\sc v} absorption,
\item Little far-UV flux between between $\lambda$200\AA\ and $\lambda$1000\AA.
\end{enumerate}
% you should describe them here.

\noindent The first criterion is simply a statement that ghost-features can
only be observed if significant photon scattering is occurring.  The second
criterion implies that there must be a significant flux of Ly-$\alpha$
photons, while the third requires that the line-emitting gas has a relatively
narrow spread in velocities so that the resulting feature is not spread over a
large range of velocity reducing the contrast. Since the dominant scattering
ion is N~{\sc v}, strong N~{\sc v} absorption is also needed. The final
requirement, and the most difficult to measure, recognises that in order for
Ly-$\alpha$--N~{\sc v} line-locking to dominate the dynamics of the flow,
there cannot be significant far-UV line driving (Korista et~al.  1993).

\section{Selection of ghost-candidate spectra}

%\onecolumn
\begin{figure}
\centering \includegraphics[width=0.5\textwidth]{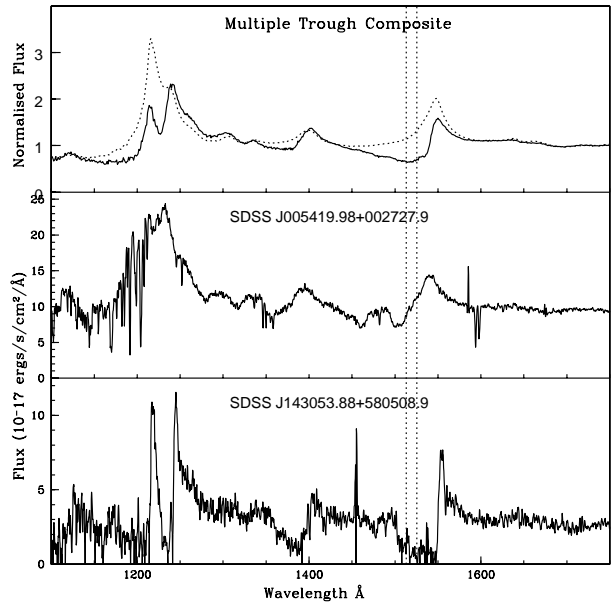}
%height=0.25\textheight]{0906MTonly}
\caption{Top-Panel: Geometric mean composite spectrum of objects
  (1019) rejected due to having multiple trough features (solid line)
  fit with a reddened DR5 QSO composite (dashed line).  Middle and
  lower panels: example spectra of individual objects rejected at this
  stage. The dashed vertical lines indicates the location of the
  ghost-zone.}
\label{fig:MT}
\end{figure}
%\twocolumn

BALQSOs represent $\sim$15\% of the quasar population (see e.g. Knigge
et~al. 2008 and references therein), of these, between $\sim$20-25\% show
evidence for multiple troughs \citep{Kor93,Nor06}.  Amongst the multi-trough
objects, ghost candidates are likely rare representing less than a few per
cent of BALQSOs (Arav 1995) because of the stringent requirements for ghost
observability.  Thus in order to find significant numbers of ghost-candidates,
large quasar samples are required.

All previous ghost candidate samples have been limited to just a handful of
objects \citep{Nor06,AravV}. By comparison, Data Release 5 (DR5) of the Sloan
Digital Sky Survey (SDSS) contains 77,429 quasars of which 28,421 have
redshifts between 1.7 and 4.2 allowing C~{\sc iv} and any associated
absorption to be seen in their optical spectra.  We require a visible C~{\sc
iv} BAL because although radiative line driving will produce a feature in all
absorption troughs, the resultant hump will be most easily observed in the
C~{\sc iv} absorption trough because of the small doublet separation of this
line.  Several BALQSO samples have been compiled from the various data
releases of the SDSS \citep{Reich03a,Trump06,Knigge} and from DR5
\citep{Gib08,Sca09}. Here we adopt the BALQSO catalogue of \cite{Sca09},
containing 3,552 BALQSOs. The Scaringi et~al. sample, differs from other
samples in that is does not rely on simple metrics such as the Balnicity index
(BI, Weymann et~al. 1991) or Absorption index (AI, Hall et~al. 2002, Trump
et~al. 2006) to identify and classify BALQSOs. Rather it uses a robust hybrid
method, involving gross classification with a supervised neural network
(learning vector quantisation) and the visual inspection of outliers.

To identify strong ghost candidates we have performed five cuts on this BALQSO
sample. The first cut eliminates noisy spectra (S/N$<$4 per pixel), estimated
in two line-free continuum bands from $\lambda\lambda$1650--1700\AA\/ and
$\lambda\lambda$1700--1750\AA.  To reduce contamination from emission,
absorption and cosmic rays that may be present in either bin we take the
greater of the two values as our S/N estimate.  The second cut removes all
objects which show an apparently single smooth absorption trough. Following
this cut we are left with only those objects which display multiple absorption
trough features (1019 in total).\footnote{Approximately 28\% (1019 out of
3552) of our BALQSOs show evidence for multiple troughs. This compares
favourably with previous studies by North et~al. (2006) who found that 58 out
of 224 (26\%) of BALQSOs from SDSS EDR showed evidence of multiple-troughs,
and Korista et~al. (1993) who found 16 out of 72 objects (22\%) showed
multiple-troughs.}

A third rejection cut removes those objects that contain 2 or more peaks
within the absorption trough (numbering 761).  These objects are eliminated to
reduce contamination by objects which display absorption from multiple
out-flowing regions which by chance happen to lie at velocities corresponding
to the ghost-zone.  This leaves just 258 objects which contain a single-peak
(or alternatively double-trough) within the C~{\sc iv} absorption region.

%\onecolumn
\begin{figure}
\centering \includegraphics[width=0.5\textwidth]{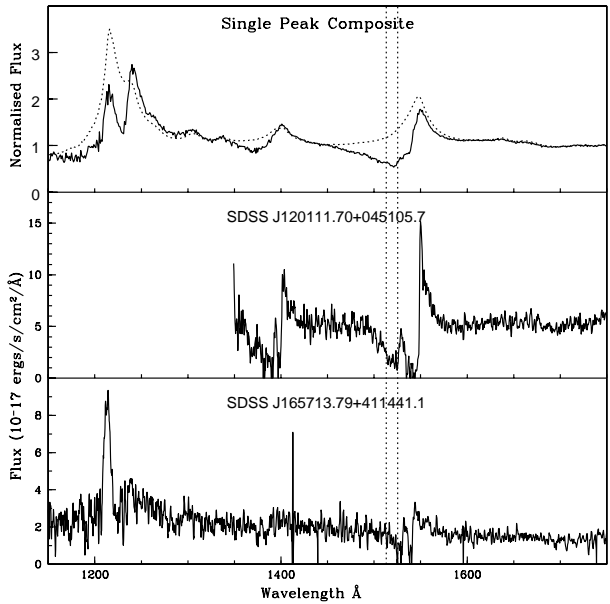}
%height=0.25\textheight]{0906SPonly}
\caption{Top-Panel: Geometric mean composite spectrum of objects (189) rejected due
to having double-trough, ghost-like features, that appear outside of the ghost
zone (solid line) fit with a reddened DR5 composite (dashed line). Middle and
lower panels: example spectra of individual objects rejected at this stage. The
dashed vertical lines indicate the location of the ghost-zone.}
\label{fig:SP}
\end{figure}
%\twocolumn

Figure~\ref{fig:MT} (upper panel) shows our multi-trough composite spectrum
(solid line) created from taking the geometric mean of all objects rejected at
this stage after normalising the flux between $\lambda\lambda$1700--1750\AA\/
to unity. In order to highlight the absorption regions, we also plot a
non-BALQSO composite spectrum (dashed line). The non-BALQSO composite is the
geometric mean of all of the quasar spectra in the SDSS DR5 quasar catalogue
in the redshift range 1.7$<$z$<$4.2 excluding those objects which are also in
the BALQSO catalogue of Scaringi et~al., and reddened to fit to the continuum
windows in the multi-trough composite spectrum using the SMC extinction law
of \cite{Pei92}. For completeness, the middle
and lower panels show two example spectra rejected at his stage.  We note that
our multi-trough composite spectrum \ref{fig:MT} shows no evidence for
coherent structure within the C~{\sc iv} trough. This suggests that the
positions of the various absorption systems in the individual spectra are in
general uncorrelated, that is they do not contain a large number of ghost
candidates that happen to show additional peaks within the absorption. This is
consistent with the hypothesis that these multiple-trough spectra are the
result of multiple unconnected absorbing systems.

%\onecolumn
\begin{figure}
\centering \includegraphics[width=0.5\textwidth]{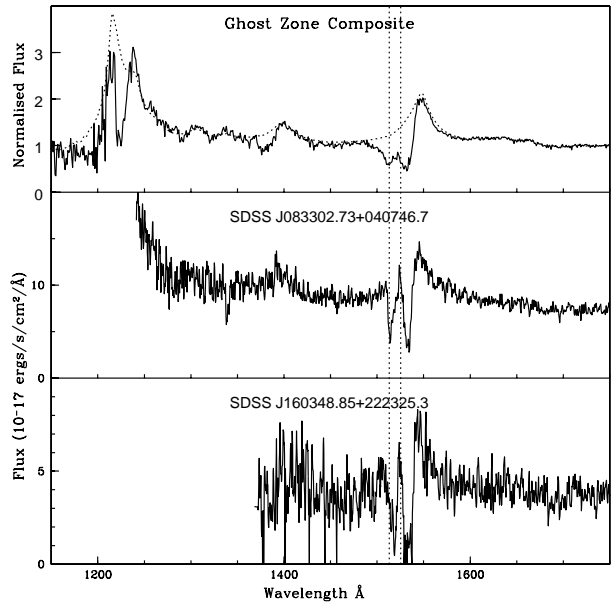}
%height=0.25\textheight]{0906GZonly}
\caption{Top-Panel: Geometric mean composite spectrum of objects (26) rejected due
to having
insufficient absorption either side of the ghost feature.  Middle and lower
panels: example spectra of individual objects rejected at this stage. The
dashed vertical lines indicate the location of the ghost-zone.}
\label{fig:GZR}
\end{figure}
%\twocolumn

For the 258 objects which show a single peak in their C~{\sc iv} absorption we
make a further cut to remove those objects (189) which show peaks outside of
the ghost-zone as defined by \cite{Nor06}. The ghost zone edges are
5900~km~s$^{-1}$ blue-ward of the peaks of the doublet of the emission line in
question (in this work C~{\sc iv}) this zone is then expanded to take into
account redshift errors by multiplying these boundaries by 1$\pm\Delta
z/(1+z_{med})$, where $z_{med}$ is the median redshift of our sample
($z_{med}=$2.14), and $\Delta z$ is the average redshift error ($\Delta
z=$0.01). This results in a ghost zone extending between $\lambda$1513.2~\AA ~
and $\lambda$1525.4~\AA ~ for C~{\sc iv}. Our single-peak composite comprising
189 objects whose peak lies outside of the ghost-zone is shown in
Figure~\ref{fig:SP} (upper panel, solid line). The middle and lower panels
show example spectra of objects rejected at this stage. Each object shows a
clear peak located just outside of the limits of the ghost zone.  
%\onecolumn
\begin{figure}
\centering \includegraphics[width=0.5\textwidth]{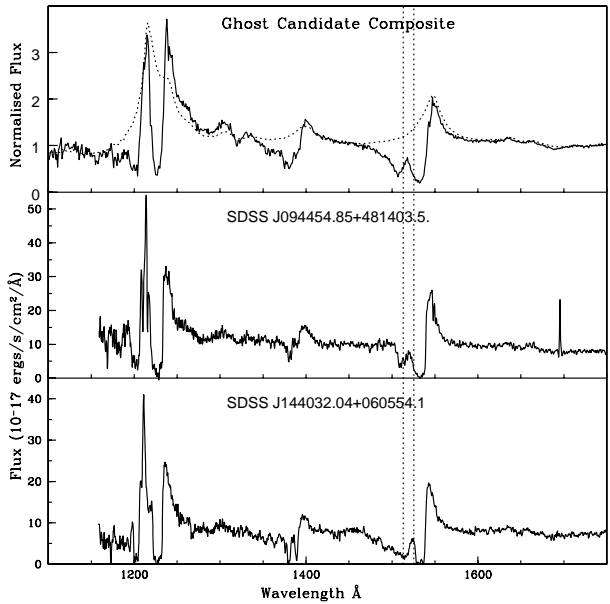}
%height=0.25\textheight]{0907GHOSTCOMP}
\caption{Top-Panel: Geometric mean composite spectrum of our ghost-candidate
  sample of 43 objects. Middle and lower
panels: example spectra of individual objects making up this sample. The
dashed vertical lines indicate the location of the ghost-zone.}
\label{fig:GZ}
\end{figure}
%\twocolumn
Interestingly, this composite spectrum shows some evidence of a feature at the
low velocity edge of the ghost zone. One possible explanation for this feature
is that these objects are genuine ghost-candidates with poorly assigned
redshifts.  However, on closer inspection of the objects rejected at this
stage we find no evidence for features outside of the ghost zone due to
random redshift errors.
Alternative explanations for the origin of this feature include, (i)
intrinsically weaker absorption at lower velocities, or (ii) an abrupt change
in the intensity of the (overlying) emission at these velocities.

\begin{figure}
\centering\includegraphics[width=0.5\textwidth,
height=0.5\textheight]{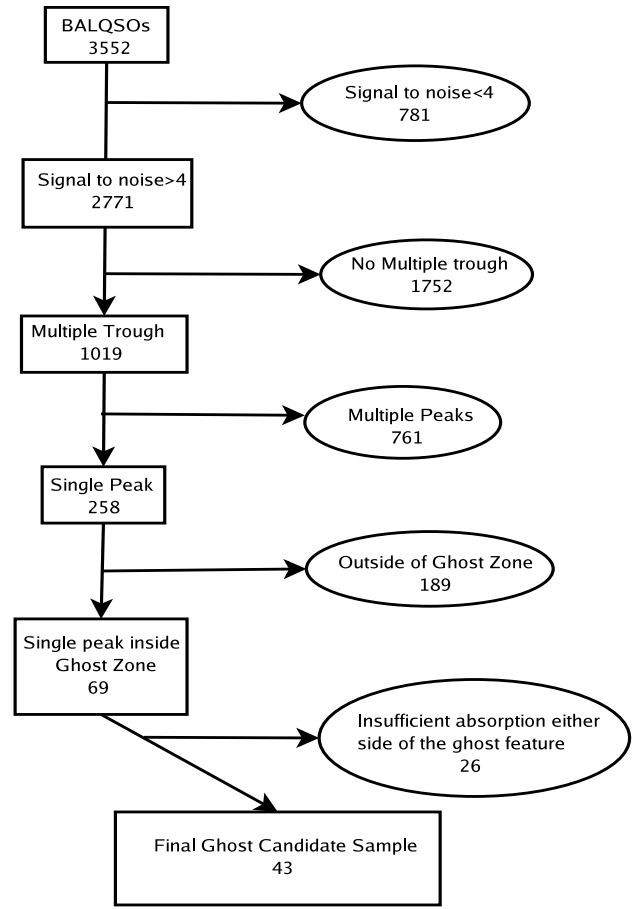}
\caption{Flow diagram showing the number of objects rejected at each step in
our ghost selection algorithm.}
\label{fig:ghostflow}
\end{figure}

We make one further cut to remove those objects which do not show significant
broad absorption either side of the ghost feature extending from 3000 to
9000~km~s$^{-1}$ blue-ward of the emission-line. This cut is chosen to
eliminate objects with deep narrow absorption features whose alignment mimics
the presence of a ghost feature.  Figure~\ref{fig:GZR} (upper panel, solid
line) displays the composite spectrum of all objects rejected at this stage,
while in the middle and lower panels, we give examples of individual objects
rejected by this cut. Though the composite spectrum shows a clear feature
within the ghost zone, it is not smooth, instead displaying sharp narrow
features, whose separation is larger than the doublet separation of C~{\sc
iv}.  This reflects both the narrowness of the absorption features (since
intrinsically broad humps would be smoothed out in the averaging process
yielding low contrast features) which make up this composite, and in addition
small differences in their velocities relative to the rest-frame of the
source. These differences may result from these features being unrelated to
the ghost mechanism and appearing in the ghost zone by chance though we cannot
preclude the fact that some of the objects rejected at this stage are indeed
true ghost-candidates.

Our final sample of 43 ghost-candidate objects comprises the largest sample of
objects exhibiting the ghost of Ly-$\alpha$ and the first to allow a
statistical analysis of the properties of these objects.  Figure~\ref{fig:GZ}
shows the geometric mean composite ghost-candidate spectrum (solid line),
together with a non-BALQSO composite spectrum (dashed line), and two example
ghost-candidate spectra (middle and lower panels).  Our final ghost-candidate
composite spectrum displays a clear smooth double-trough structure located
firmly within the ghost zone of a relatively broad absorption trough.  Visual
comparison of this composite with that formed from objects rejected at the
previous stage (i.e. those with insufficient absorption either side of the
ghost-zone), indicates that our final sample displays on average narrower
Ly-$\alpha$ and C~{\sc iv} emission-lines, stronger Ly-$\alpha$ emission and
deeper N~{\sc v} absorption (fulfilling 3 of Arav's criteria for forming
observable ghost features), as well as broader Si~{\sc iv} absorption, and
suggests that we are indeed isolating those objects most likely to form
observable ghost features.  Figure~\ref{fig:ghostflow} summarises the various
cuts made in the creation of our ghost-candidate sample.

\subsection{Selecting a Comparison Sample}

\begin{figure}
\centering\includegraphics[width=0.5\textwidth,
height=0.5\textheight]{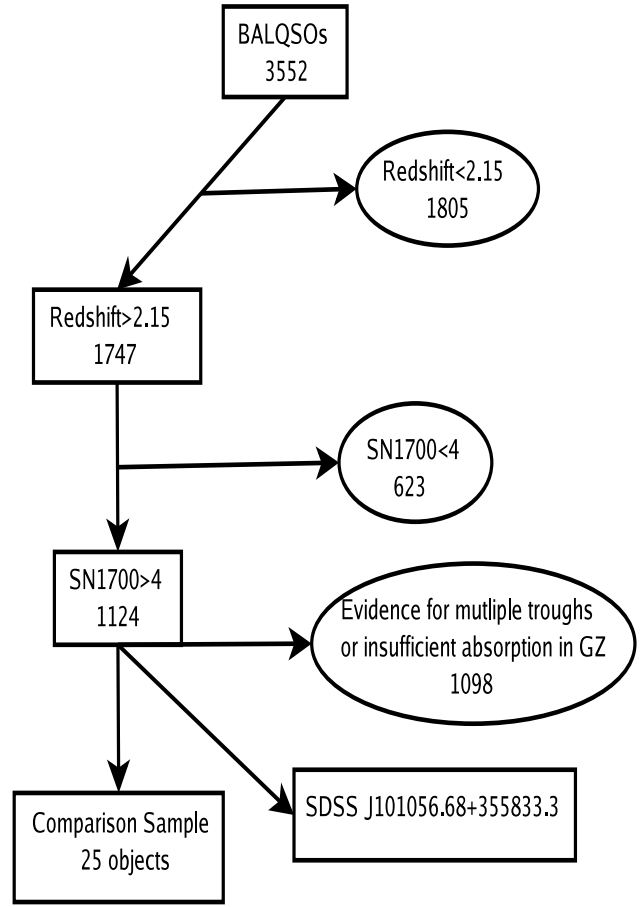}
\caption{Flow diagram showing the number of objects rejected at each step in
our comparison sample selection algorithm.}
\label{fig:comparison_flow}
\end{figure}

Of the 43 strong ghost-candidates, 21 are at large enough redshift $z> 2.15$
to place Ly-$\alpha$ above the atmospheric cut-off and thus allow us to test
for each object all of the criteria deemed necessary for
ghost-formation. These are listed in the upper half of Table~1.  For the
remainder, their redshifts are too low to allow a direct measurement of the
strength and width of the Ly-$\alpha$ emission-line. For these objects we
estimate the strength and width of the Ly-$\alpha$ emission-line using the
C~{\sc iv} emission-line as a surrogate, and substantiated by known
correlations between the two lines (see e.g. Wilkes 1984, Ulrich 1989). These
objects are listed in the lower half of Table~1.  For completeness, we have
also compiled a sample of BALQSOs showing no evidence for a ghost feature or
multiple trough. This allows us to compare the relative frequency with which
the ghost-formation criteria are met amongst BALQSOs both with and without
ghost features.  In order to create such a sample we have performed a number
of cuts on the \cite{Sca09} BALQSO sample. As for our ghost-candidate sample
we first select for S/N$>$4, and then select those objects with $z>2.15$ to
allow inspection of the N~{\sc v} and Ly-$\alpha$ emission-lines necessary to
test for the presence of strong narrow Ly-$\alpha$ emission and strong N~{\sc
  v} absorption.  The remaining spectra are then visually inspected and any
object showing multiple absorption troughs or any other hint of a ghost
feature are removed from the sample. We also remove those objects which do not
show significant absorption between 3000 and 9000~km~s$^{-1}$.  This leaves
just 26 objects showing significant C~{\sc iv} absorption between 3000 and
9000~km~s$^{-1}$ with no evidence for a ghost feature or multiple trough. One
of these objects SDSS J101056.68+355833.3 shows evidence for a ghost feature
in the Si~{\sc iv} absorption line (see A5) and is therefore also removed from
our comparison sample leaving just 25 objects.  A summary of the rejection
cuts made in the creation of our comparison sample is given in
Figure~\ref{fig:comparison_flow}. The geometric mean composite spectrum of our
comparison sample is indicated by the dotted line in Figure~7.  Aside from
substantially weaker N~{\sc V} absorption, the strengths and widths of the
emission-lines in the comparison sample composite spectrum match those of the
ghost candidate composite spectrum remarkably well. The main difference
between these two composite spectra appears to be the absence of the
low-velocity absorption trough in the comparison sample spectrum. We do not
believe this difference arises from the way in which the two samples were
selected.

In \S4 we use our ghost-candidate sample to test, where possible, the criteria
set out by \cite{AravV} required by radiative acceleration models to produce
an observable ghost feature. We compare the relative frequency by which our
ghost candidate spectra meet the criteria necessary for the formation of
observable ghost features with our non-ghost comparison sample.

\section{Testing the criteria for the formation of observable ghost features}

In this section we present the results of testing the criteria for the
formation of observable ghost features.  We first concentrate on the sample of
21 ghost-candidate objects for which all of the criteria may be tested and
compare these results with those obtained using our non-ghost comparison
sample. We then move on to our remaining objects and test where possible
whether these objects satisfy the necessary criteria.
 
\subsection{Significant C~{\sc iv} broad absorption line}

All of the objects in our ghost candidate sample will clearly show significant
absorption in the region between 3000 and 9000~km~s$^{-1}$ blue-ward of the
C~{\sc iv} emission line as this is required in order to see any potential
ghost feature. This criterion is required purely because in order to effect the
out-flowing C~{\sc iv} ions in an observable way this outflow must reach
significant optical depth and appear as a BAL trough within the spectrum.
Our comparison sample has also been chosen to exhibit strong broad absorption,
so naturally also fulfils this criterion.

\subsection{Strong Intrinsic Ly-$\alpha$ emission}
\begin{figure}
\centering\includegraphics[width=0.5\textwidth]{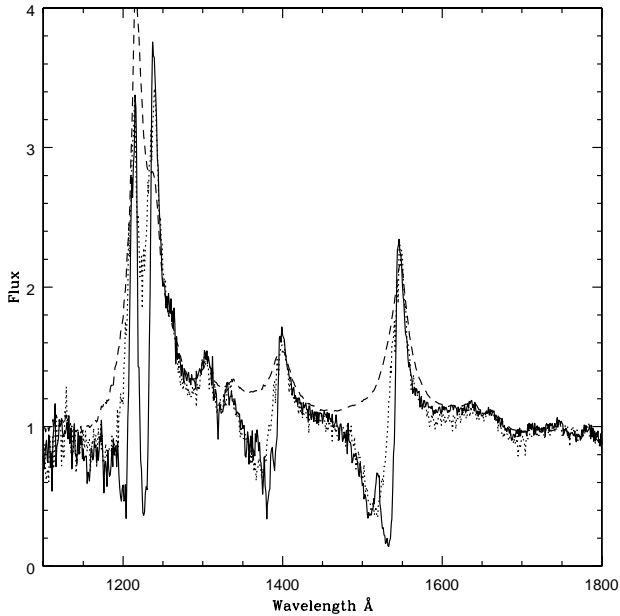}
%height=0.5\textheight]{090713_nonBAL_ghost_comparison}
\caption{Comparison between the geometric mean composites formed from (i) 21
  ghost-candidates with $z>2.15$ (solid line), (ii) 25 BALs with featureless
  broad absorption troughs (dotted line), and (iii) non-BALQSOs from the DR5
QSO catalogue (dashed line).}
\label{fig:comp_samples}
\end{figure}

In order for the effects of Ly-$\alpha$ radiation on N~{\sc v} to cause a
significant increase in the radiation pressure on the outflow as a whole the
Ly-$\alpha$ emission line must be intrinsically strong.  However, since the
ghost of Ly-$\alpha$ is produced by scattering Ly-$\alpha$ photons it can be
extremely difficult to measure the strength of the intrinsic Ly-$\alpha$
emission-line. Hence \cite{AravNature} modified this criterion to allow the
C~{\sc iv} and N~{\sc v} emission-lines to be used in situations where the
Ly-$\alpha$ emission-line was either significantly absorbed, or had
insufficient spectral coverage. This modified criterion requires Ly-$\alpha$
emission to be at least 300\% above the continuum and/or both C~{\sc iv} and
N~{\sc v} to be at least 100\% above the continuum\footnote{We note that
measuring the strength of a line according to its peak flux can be misleading,
since it implies that the underlying line profiles are similar for all objects
(and thus the total flux in the line simply scales with the peak flux).  For
example, a comparison between the C~{\sc iv} emission-line of the composite
spectra in Figure~7 suggests that both our ghost-candidate and comparison
samples display a narrower underlying C~{\sc iv} emission-line. If the widths
of the other emission-lines scale with the C~{\sc iv} emission-line width,
then we would expect both the Ly-$\alpha$ and N~{\sc v} emission-lines to be
similarly narrow.}  .  They note that `the latter criterion is hardly ideal but
observations of non-BAL quasars show that the line strengths roughly scale
together' (Arav et~al. 1995). Here we isolate the continuum emission using the
{\sc specfit} package in {\sc iraf}. We model the continuum as a reddened
power law and fit to the continuum in regions free of contaminating emission-
and absorption- lines, where possible. This can be particularly difficult for
objects with the highest velocity out-flows as there are few
emission/absorption line-free regions.  Furthermore, we have not attempted to
fit for the continuum short-ward of $\lambda$1280\AA\/ as this region is
typically so absorbed and the line-emission so highly blended that no clean
continuum regions can be identified. Instead, we fix the continuum strength at
short wavelengths to that measured between $\lambda\lambda$1315--1350\AA.
This approach is chosen to avoid overestimating the strength of the broad
absorption, at the expense of overestimating the strength of the line-emission
in a few cases.

In objects whose spectra cover the Ly-$\alpha$ region, 86\% (18/21) of the
ghost candidate sample meet the criterion for strong emission-lines compared
to 88\% (22/25) of the comparison sample.  Inspection of the composite spectra
for the two samples (Figure~\ref{fig:comp_samples}), indicates that the
strength of the Ly-$\alpha$ emission-line (relative to the continuum) is
broadly similar in the two samples, and both are weaker relative to the
non-BALQSO composite, due to the presence of strong N~{\sc v} and Ly-$\alpha$
absorption.  For the other lines, the N~{\sc v} emission-line appears stronger
on average in our ghost-candidate and comparison samples relative to
non-BALQSOs, while there appears little difference in the strength of the
C~{\sc iv} emission-line, though we note that in so far as we can measure
them, their emission-line widths appear to be narrower on average.

\subsection{Narrow emission-lines}

Models of outflows accelerated by radiative line driving produced by
\cite{AravIII,AravNature} suggest that once the line widths exceed
3500~km~s$^{-1}$ (FWHM) potential ghost features become undetectable.  This is
because broader emission-lines will result in a broader, lower contrast
feature (since the core of the line, where most of the scattering occurs, is
weaker relative to the wings in a broad emission-line c.f a narrow
emission-line) which is difficult to detect particularly in low S/N
data. Unfortunately, determining the width of the underlying emission-lines in
BALQSOs is extremely difficult as the broad absorption in these objects
prevents accurate measurements of the shape and strength of the underlying
emission-lines and continuum.  Given how our ghost-candidate sample was
selected we expect all of the ghost-candidates to show significant absorption
in the blue wing of Ly-$\alpha$, N~{\sc v} and C~{\sc iv}, while our
comparison sample was chosen to exhibit strong relatively featureless
absorption in the blue wing of C~{\sc iv}.

Since we cannot make an accurate measurement of the width of the Ly-$\alpha$
emission-line we instead attempt to estimate its likely width using the C~{\sc
iv} emission-line width as a surrogate.  For the majority of sources, C~{\sc
iv} is severely absorbed, so we estimate its' underlying un-absorbed width
using model fits to the red wing of the emission-line.  We use the same
underlying fit for the continuum as was used to measure the emission-line
strength (\S4.2).  We fit the red-wing with a two-component model, a Gaussian
core and Lorentzian wings, with the central wavelengths fixed together but
allowed to vary slightly from the rest wavelength of C~{\sc iv}. We allow the
strength and width of the two components to vary independently and fit to the
red-wing taking care to avoid those regions of the spectrum contaminated by
broad He~{\sc ii}~$\lambda$1640 and [O~{\sc iii}] ~$\lambda$1663\ emission.
The measured width is taken to be the FWHM of the composite fit.  If the
measured FWHM$<$3500~km~s$^{-1}$, then the object meets the criterion for
the formation of an observable ghost feature.

For our 21 ghost-candidates with z$>$2.15, only 12 (57\%) have FWHM (C~{\sc
iv}) $<$~3500~km~s$^{-1}$. In the remainder, those for which the C~{\sc iv}
width can actually be measured (17 objects), 8/17 (47\%) also meet this
criterion. This fraction is similar to that found for our comparison sample
where 12 out of the 20 objects (60\%) for which C~{\sc iv} width measurements
can be made, also meet this criterion (see Table~2 for details).  Taking the
average of all of the FWHM measured for the ghost candidates gives an average
FWHM of 3345$\pm$842~km~s$^{-1}$ which is similar to the upper limit proposed
by \cite{AravV} and to the average FWHM of C~{\sc iv} found for the comparison
sample (3534$\pm$1160~km~s$^{-1}$). A K-S test on these populations shows no
significant difference between the ghost and comparison samples. The
difficulty in fitting the C~{\sc iv} line in many of these objects results in
large uncertainties in the estimated widths, however it is clear that several
of our ghost candidates have emission that is wider than the 3500~km~s$^{-1}$
upper limit proposed by \cite{AravV}. There are a number of possible
explanations for this.  Firstly, C~{\sc iv} may be a relatively poor surrogate
for the width of Ly-$\alpha$. Alternatively, our sample may suffer from
contamination by objects in which the potential ghost feature is produced by a
chance alignment of multiple absorption systems or through other mechanisms
unrelated to radiative acceleration.  If the potential ghost-features are
indeed due to line-locking, then it may in-fact be possible to produce an
observable ghost feature from emission-lines with widths in excess of
3500~km~s$^{-1}$.

\subsection{Strong N~{\sc v} broad absorption line}

As described in \S2, the ghost feature is produced by the increased radiation
pressure on out-flowing N~{\sc v} ions by Ly-$\alpha$ photons.  Since in this
model the N~{\sc v} ions scatter a significant fraction of the incident flux,
strong N~{\sc v} absorption is a pre-requisite.
 
In order to measure the strength of absorption due to N~{\sc v} we use a
modified Balnicity index (Weymann 1991).  We take the continuum to be equal to
the mean flux between $\lambda\lambda$1315--1330\AA, and require that the flux
drops to below 90\% of this value continuously for $>1000$~km~s$^{-1}$ between
0 and 7000~km~s$^{-1}$ blue-ward of the N~{\sc v} emission line. We use the
7000~km~s$^{-1}$ limit to avoid contamination by absorption due to
Ly-$\alpha$. Any object with a non-zero value for this measure is considered
to show strong N~{\sc v} broad absorption.

Significant N~{\sc v} absorption is measured in all 21 objects from our ghost
candidate sample whose redshift allows detection of N~{\sc v} absorption.
Only 44\% (11/25) of our comparison sample satisfy this criterion. The
geometric composites of Figure~7, confirm that our ghost-candidate sample
displays significantly stronger N~{\sc v} absorption on average than our
comparison sample. Moreover, the broadly similar N~{\sc v} emission-line
strengths of the ghost-candidate and comparison sample, emphasises the
increased importance of N~{\sc v} absorption in the ghost-candidate spectra.

\subsection{Little far-UV flux}
Photo-ionisation models suggest that the far UV spectra of quasars contain a
significant number of emission-lines between $\lambda$200 and
$\lambda$1000\AA\/ that will contribute to the radiation pressure on any BAL
outflow (Korista et~al. 1993). In order for a ghost feature to be observed the
radiation pressure of Ly-$\alpha$ on the N~{\sc v} ions in the outflow must be
at least comparable in strength to the radiation pressure on all other ions in
the outflow.  This led Arav to suggest that the far-UV continuum must
therefore be necessarily weak in objects showing ghost features.  Absorption
of far-UV photons both within the host galaxy as well as within our own galaxy
preclude direct measurement of the far-UV continuum. However \cite{AravNature}
note that the strength of the He~{\sc ii}~$\lambda1640$~\AA ~emission-line can
be used as a surrogate for the far-UV ionising continuum. A strong He~{\sc ii}
line suggests a similarly strong far-UV flux and vice-versa as this line is
produced by photons with energies corresponding to $\lambda$228\AA
($\approx$54~eV).

However, the He~{\sc ii} emission-line strength is notoriously difficult to
measure even in high quality spectra. This is because of its close proximity
to the [O~{\sc iii}]$\lambda$1663 emission-line, as well as its location near
the red-wing of C~{\sc iv}, which may extend to relatively high velocities.
Thus isolating the broad He~{\sc ii} component generally requires
multi-component fitting (see e.g. Goad and Koratkar 1998) which without prior
information (for example, variability data), is rather subjective.

\begin{figure}
\centering\includegraphics[width=0.5\textwidth]{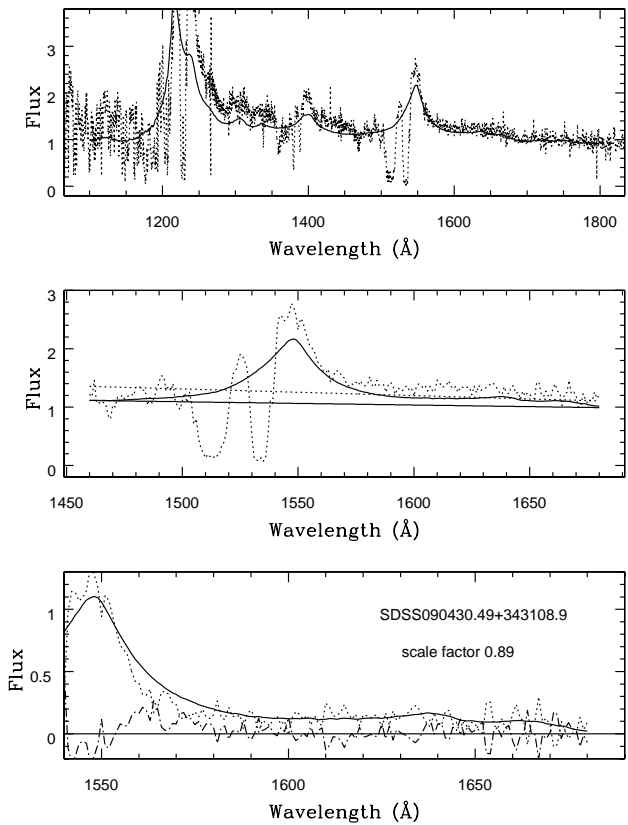}
%height=0.5\textheight]{plot_37}
\caption{An illustration of our method for estimating the He~{\sc ii}
strength. Upper panel -- non-BALQSO composite (solid line) and an example
ghost-candidate spectrum (dashed line). Middle panel -- model fits to the
underlying continua. Lower panel -- non-BALQSO composite (solid line) and
scaled ghost-candidate spectrum, after subtraction of the underlying
continuum. The residual flux is indicated by the dot-dashed line.}
\label{fig:heii_fit}
\end{figure}

Since the quality (in terms of S/N) of the individual spectra in both our
ghost-candidate and comparison samples is relatively low, we do not attempt to
fit to the individual line components. Instead, we use our non-BALQSO
composite spectrum as a template for the region of interest, and scale each
spectrum to this template, by minimising the residual flux in the region of
interest. Our optimisation routine fits for the underlying continuum in both
spectra, and adjusts for both redshift errors and the scale factor over the
region of interest (nominally $\lambda\lambda$1600-1680\AA\/).  Since both
spectra are first normalised to the flux in the continuum band between
$\lambda\lambda$1700--1750\AA, if we assume that the He~{\sc ii} and [O~{\sc
iii}] emission line strengths scale together, then the derived scaling factor
between the ghost-candidate and non-BAL QSO spectra is a measure of the ratio
of their He~{\sc ii} EWs.  Figure~\ref{fig:heii_fit} shows an example of a
fit.  In the top panel we show the non-BALQSO composite (solid line) and a
ghost-candidate spectrum (dotted line). In the middle panel we show the
spectra together with their continuum fits spanning the C~{\sc iv}--He~{\sc
ii} region, while in the lower panel we show a close-up of the non-BALQSO
composite and the scaled ghost-candidate spectrum after removal of the
underlying continuum. The residual flux under the He~{\sc ii}--[O~{\sc iii}]
lines is indicated by the dot-dashed line. We caution that this fitting
process is sensitive to (i) small redshift errors between the spectra, (ii) an
appropriate choice of continuum bands (in general the choice of the continuum
bands depend on the width of the C~{\sc iv} absorption trough, and most
importantly (iii) the S/N of the spectra.  Consequently the error on the scale
factor is generally quite large.  We have visually inspected the results of
the fitting process for all of our ghost-candidate and comparison spectra to
verify that the fitting procedure has converged correctly, adjusting the
continuum bands, and region over which the spectra are scaled when
necessary. In Table~1 we list the He~{\sc ii} scale factors, that is, the
multiplication factor necessary to produce the same equivalent width in the
ghost-candidate spectrum. A number less than 1 indicates that the
ghost-candidate spectrum has larger equivalent width in the line in the region
of He~{\sc ii} than the non-BALQSO spectrum. Conversely, numbers greater than
1 indicate that the ghost feature has a proportionately weaker He~{\sc ii}
line.  In order to determine error estimates on the derived scale factors,
after determining a best-fit solution, we fix all parameters apart from the
scale factor, and then minimise on this parameter only. We then calculate the
90\% confidence interval on this one interesting parameter from the best-fit
model by varying the scale factor until the $\chi^2$ value has increased by
2.71. Both the best-fit scale factor and its range (based on the 90\%
confidence interval), are given in Table~1 and 2.

For our sample of 21 ghost-candidates for which all criteria can be
tested, 5 indicate a significantly weaker He~{\sc ii} EW, suggesting a
weaker than average UV continuum. Only 5 objects show evidence for a
stronger than average He~{\sc ii} EW. For the rest, there is no
significant difference (within the errors) in their He~{\sc ii} EWs
when compared to the non-BALQSO composite.  For the 22 ghost-candidates
for which only some of the criteria can be tested, 5 show evidence for
a weaker than average He~{\sc ii} EWs, 6 indicate stronger than
average He~{\sc ii} EWs, and the rest indicate no significant
difference in the He~{\sc ii} EW when compared to the non-BALQSO
composite. Thus only 11/43 (25\%) of our ghost-candidate sample appear
to satisfy the criteria for a weak UV continuum.  Repeating this test
on our non-ghost BALQSO comparison sample 8 objects indicate weaker
than average He~{\sc ii} EW, 9 objects have stronger than average
He~{\sc ii} EW, and 6 objects show no discernible difference in their
He~{\sc ii} EW relative to the non-BALQSO composite. For the remaining
2 objects, the fits did not converge, though visual inspection of one
of these shows no evidence for significant He~{\sc ii} emission.

\subsection{Supplementary requirements - the width of the ghost feature}
% depends on optical depth effects, may only scatter from the core of
% the line

\begin{figure}
%\onecolumn
\centering \includegraphics[width=0.5\textwidth]{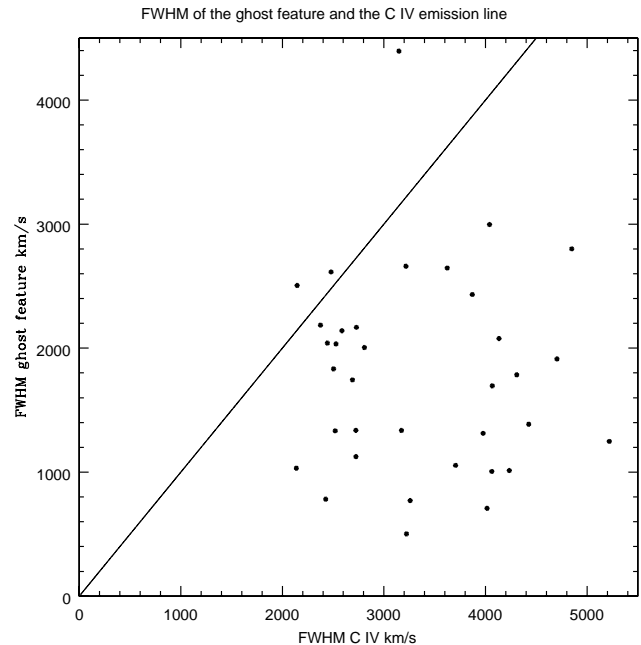}
%height=0.38\textheight]{0906FWHM}
\caption{FWHM (km~s$^{-1}$) of the C~{\sc iv} emission-line and the ghost
feature. The solid line shows FWHM(C~{\sc iv})$=$FWHM(ghost).}
\label{fig:FWHM}
\end{figure}

If we ignore the likely complex interaction between the Ly-$\alpha$ photons
and N{\sc v} ions (ie. we assume the optical depth in the flow is effectively
constant with velocity), then naively one might expect that the ghost-feature
should be at least as broad as the Ly-$\alpha$ emission-line responsible for
its formation. We note that in fact scattering is likely dominated by the line
core where the optical depth is larger.  By requiring that any observed ghost
feature is at least as broad as Ly-$\alpha$, we can further refine our
ghost-candidate sample to leave only the best ghost-candidates.
While the width of any ghost-feature is relatively easy to measure (unless the
S/N is low), we are once again limited by the accuracy to which we can 
measure the width of Ly-$\alpha$, as this emission is absorbed in
the process by which the ghost is formed. Here we again make use of the width
of the C~{\sc iv} emission-line (see \S4.3) as a surrogate for the width of
Ly-$\alpha$.
We isolate the ghost feature by removing the underlying absorption trough by
fitting a simple function (cubic spline). We then fit a single Gaussian to the
ghost feature to provide an estimate of its width.  The measured widths are
listed in Table~1.  We find no correlation between the width of the ghost
feature and the width of the C~{\sc iv} emission-line (Figure~\ref{fig:FWHM}).
Instead, we find that in the vast majority of cases the ghost feature is
narrower than the C~{\sc iv} emission-line (the solid line in
Figure~\ref{fig:FWHM} is FWHM(C~{\sc iv})$=$FWHM(ghost). 

If we require that the FWHM of the ghost feature matches that of the C~{\sc
iv} emission-line to within 1500~km~s$^{-1}$ then 11/20 (for one object we
are unable to make emission-line width measurements) of our ghost-candidates
for which all ghost-formation criteria can be tested also match this criterion.
For our remaining ghost-candidates 8/17 (5 objects have no emission-line width
information) also match this criterion. Of the 11 candidates with similar ghost
 and C~{\sc iv} widths for which all the criteria are testable all 11 show 
N~{\sc v} absorption, 10 show strong emission lines, 9 have narrow C~{\sc iv}
 emission but only 2 show evidence for weaker than average He~{\sc ii} 
emission. Of these two one has C~{\sc iv} FWHM of 4040~km~s$^{-1}$ while 
the other fails the criteria for strong emission lines.
%\twocolumn

\section{Discussion}

For the 43 objects in our ghost candidate sample, 21 are at large enough
redshift to allow us to test {\em all\/} of the criteria necessary for the
formation of an observable ghost feature.  Of these 21/21 satisfy the condition
for strong N~{\sc v} absorption. 18/21 also satisfy the condition for strong
intrinsic Ly-$\alpha$ emission. However, less than half of the objects (12/21)
satisfy the condition for narrow emission-lines, and even fewer (5/21) show
weaker than average He~{\sc ii} EW. While 11/21 objects satisfy the first 3
criteria, no single object meets all of the requirements for the formation of
an observable ghost feature.  For the remaining 22 objects, only 3 objects
pass 2 of the first 3 criteria, and of these all show stronger than average
He~{\sc ii}. Of the 5 objects which indicate weaker than average He~{\sc ii}
strengths, only one can be tested for any of the other criteria, in this case,
the width of the emission-line, which it fails.

None of the objects in our ghost-candidate sample are present in the samples
produced by \cite{AravV} or \cite{Kor93} due to differences in the redshift
ranges over which they were constructed.  North et~al. (2006) identified 7
strong ghost-candidate spectra in DR3 of the SDSS.  6 of those objects are
also in our sample of 43 ghost-candidates. The other, SDSSJ142050.34-002553.1.
has an assigned redshift of 2.085 in DR5 compared to the 2.103 used by
\cite{Nor06} and is therefore rejected by our ghost zone cut. Only one of
North et~al.'s ghost zone final cut is of sufficient redshift to test all of
the criteria for the formation of an observable ghost feature SDSS
J110623.52-004326.0. This object fails the narrow emission-line
requirement. For the other 5 objects, 3 fail the test for narrow
emission-lines, and 4 fail the test for weaker than average He~{\sc ii} EW.

For our non-ghost comparison sample 12/25 meet the criterion for strong N~{\sc
v} absorption.  20/25 meet the requirement of strong Ly~$\alpha$, while 12/22
meet the narrow emission-line width constraint. 9/25 objects show evidence for
weaker than average He~{\sc ii} strengths. However, only 3 objects meet 3 out
of 4 criteria, and none of these meet the weaker than average He~{\sc ii}
strength. Thus, {\em none\/} of the objects in our comparison sample meet all
of the criteria necessary for the formation of an observable ghost feature.

A comparison between our ghost-candidate and non-ghost samples suggests that
the main difference between them lies in the strength of the absorption, with
our ghost candidate sample displaying more objects with strong N~{\sc v}
absorption. Comparison of the geometric mean composite spectra of these two
samples, and their ratio (Figure~13 dashed (ghost-candidate) and dotted
(non-ghost comparison spectra) lines, indicates that the ghost-candidate
composite shows significantly stronger absorption at lower velocities in all
of the strong lines Ly-$\alpha$, N~{\sc v}, Si~{\sc iv} and C~{\sc iv}.

In order to test whether peaks are more common within the ghost-zone than
elsewhere, we repeat the ghost selection method using two ``fake'' ghost-zones
(red-ward and blue-ward of the original ghost-zone), in a similar fashion to
North et~al. (2006).  These zones are created in precisely the same way as the
ghost-zone except that the red zone is centred at 4000~km~s$^{-1}$ while the
blue zone is centred at 8000~km~s$^{-1}$ blue-ward of the C~{\sc iv}
emission-line.  Of the 258 single peaked objects, 82 have peaks within the
``fake'' red zone and 50 have peaks within the fake ``blue-zone''. Since the
original ghost-zone contained 69 objects with single peaks, the evidence for
an excess of objects with peaks at a preferred velocity is weak. That is,
single peaks within the ghost zone are no more likely than single peaks at
other velocities.

We have also examined the link, if any, between the peaks within the C~{\sc
iv} absorption trough and N~{\sc v} absorption. In order to select N~{\sc v}
BALs we use a modified Balnicity index (Weymann 1991).  We take the continuum
to be equal to the mean flux between $\lambda\lambda$1315--1330\AA, and
require that the flux drops to below 90\% of this value continuously for
$>1000$~km~s$^{-1}$ between 0 and 7000~km~s$^{-1}$ blue-ward of the N~{\sc v}
emission line.  In order to perform this test we require objects with
redshifts in excess of 2.15.  From a sample of 1747 objects with z$>$2.15, we
find 1258 N~{\sc v} BALs and 489 N~{\sc v} non-BALs.  27\% (340) of the N~{\sc
v} BALs show multiple troughs in their C~{\sc iv} absorption compared to only
15.5\% (76) of the N~{\sc v} non-BALs.  Similarly, 7.6\% (95) of the N~{\sc v}
BALs and only 3.1\% (15) of the N~{\sc v} non-BALs show a single peak in the
C~{\sc iv} absorption trough. Among the N~{\sc v} non-BALs we find : i) two
objects with single peaks within the ghost-zone, ii) two objects with a single
peak in the fake blue zone, and iii) no objects with a single peak in the fake
red zone.  For the N~{\sc v} BALs, we find : i) 26 objects with single peaks
within the ghost-zone, ii) 14 with single peaks within the fake blue zone, and
iii) 33 with single peaks within the fake red zone. While these results
indicate a strong link between the presence of N~{\sc v} absorption and the
mechanism responsible for producing features within the C~{\sc iv} absorption
trough, line-locking between Ly-$\alpha$ and NV does not appear to be the
dominant mechanism.

In summary, N~{\sc v} BALs are more likely to have multiple troughs within the
C~{\sc iv} absorption than N~{\sc v} non-BALs (factor of 2). Further, N~{\sc
v} BALs are also more likely to display single-peaks within their C~{\sc iv}
absorption than N~{\sc v} non-BALs. Approximately 25\% of the single peaked
objects are located within the ghost-zone.  However, similar numbers are found
in both the blue and red fake ghost zones. Thus while strong N~{\sc v}
absorption appears to be a strong requirement for the appearance of features
within the C~{\sc iv} absorption trough of BALQSOs, there is no preferred
velocity for the location of these features.

%\onecolumn
\begin{figure}
\centering \includegraphics[width=0.5\textwidth]{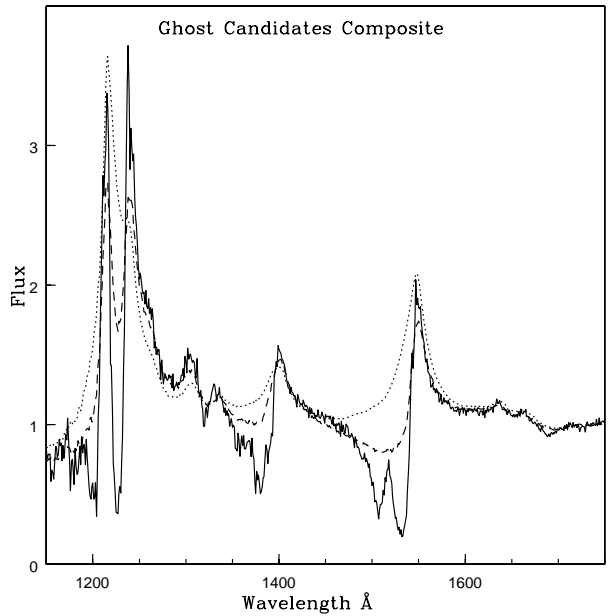}
\caption{Composite spectrum made up of all our ghost candidates fit with a
reddened DR5 composite(dotted) and a BALQSO composite (dashed).}
\label{fig:FC}
\end{figure}
%\twocolumn

\onecolumn
\begin{figure}
\centering \includegraphics[width=0.95\textwidth]{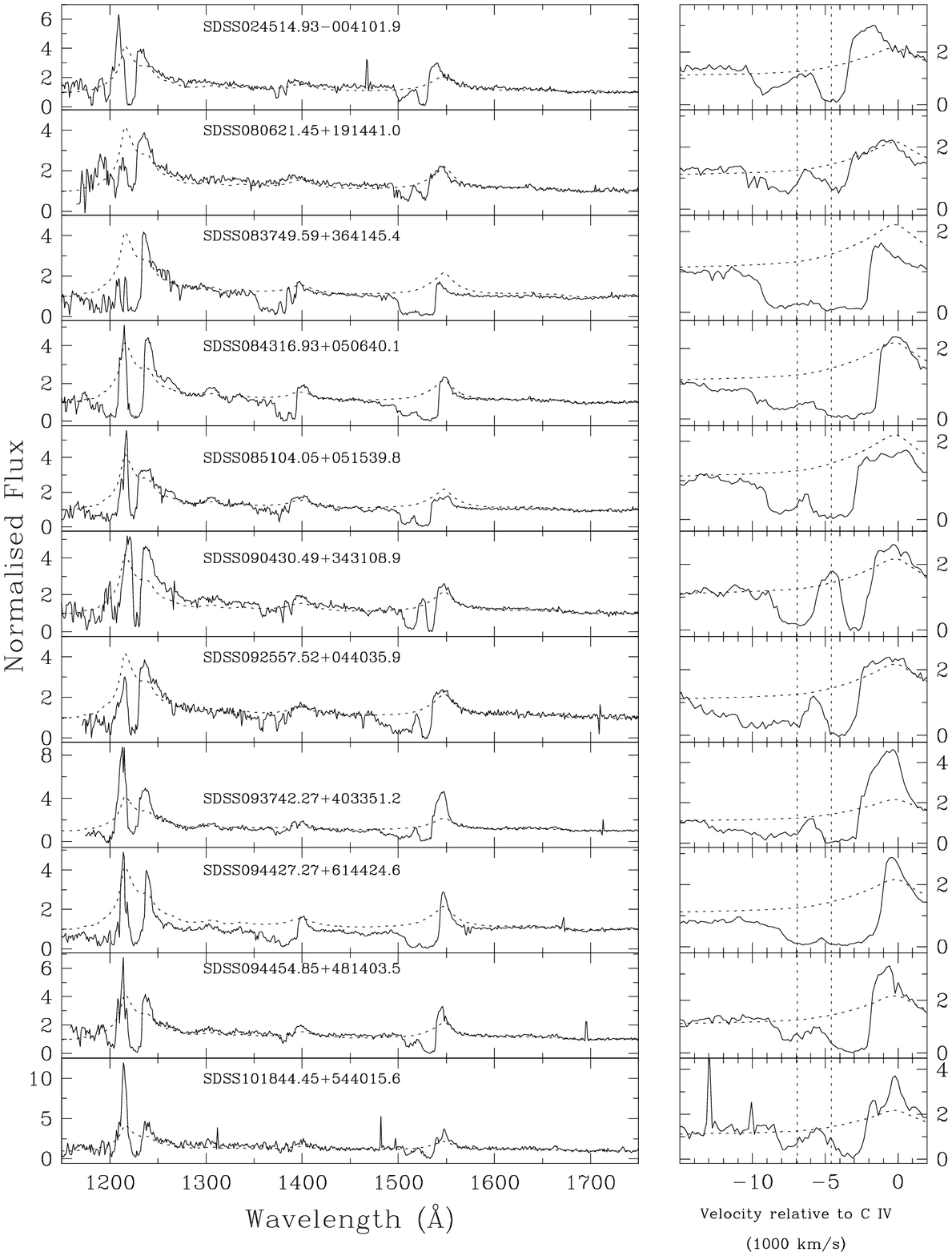}
\caption{Spectra of ghost candidates from the best-candidate sample for which
  all criteria can be tested.}
\label{fig:mike_1b}
\end{figure}
\twocolumn

\onecolumn
\begin{figure}
\centering \includegraphics[width=0.95\textwidth]{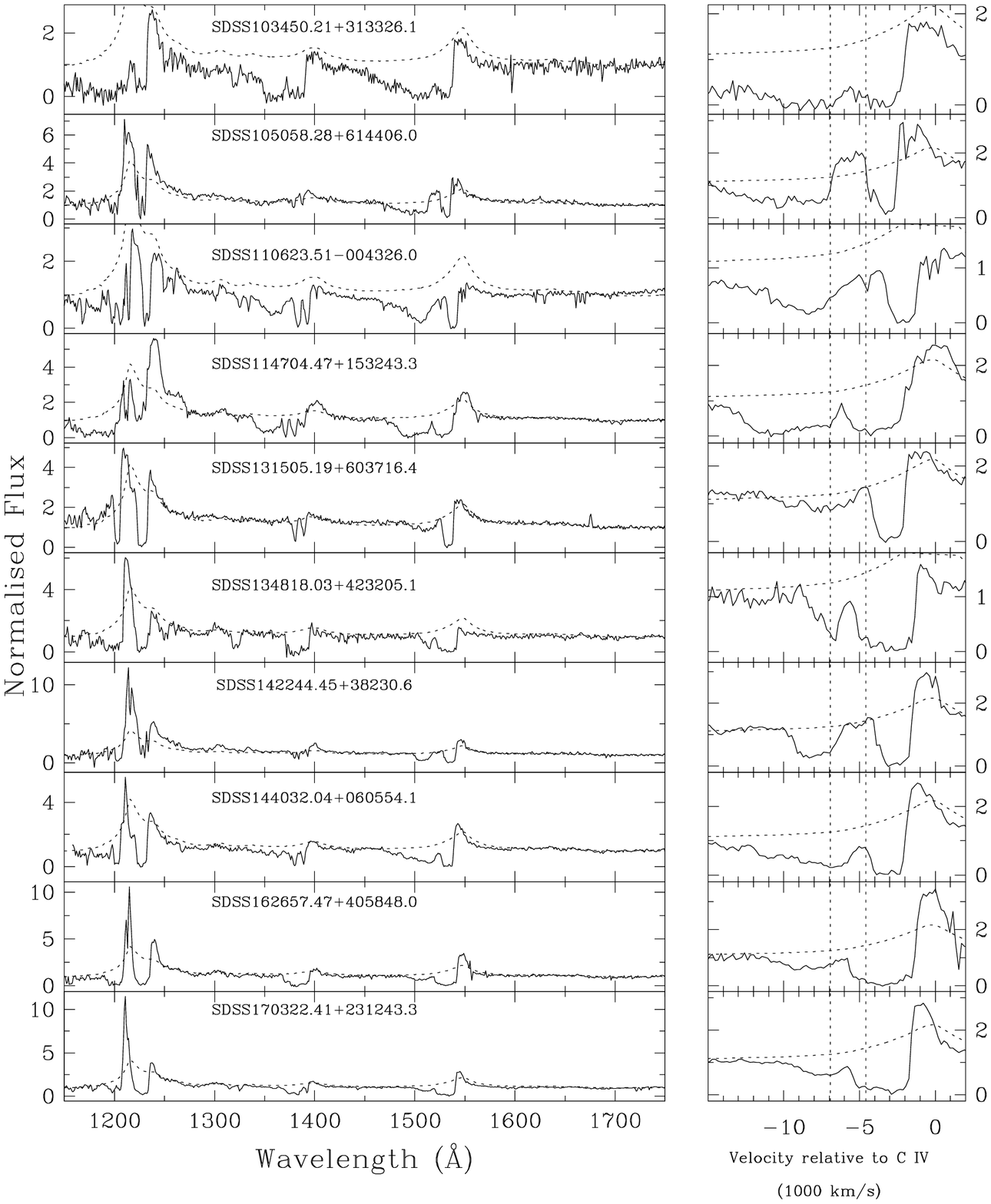}
\caption{Spectra of ghost candidates from the best-candidate sample for which
  all criteria can be tested.}
\label{fig:mike_2b}
\end{figure}
\twocolumn

\begin{figure}
\centering \includegraphics[width=0.5\textwidth]{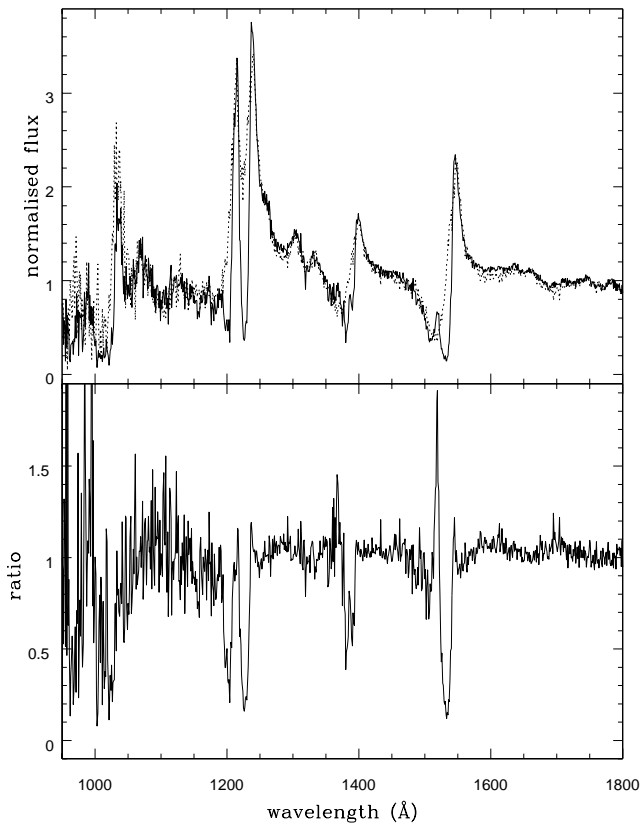}
\caption{Upper-panel - composite ghost-candidate (solid line) and comparison
  (dashed line) spectra. Lower panel - ratio of ghost-candidate to comparison spectra.}
\label{fig:plot_cand}
\end{figure}

\section{Summary of Results}

In an analysis of the largest sample of BALQSOs derived from DR5 of SDSS, we
have identified 43 objects which show evidence for radiative driving, as
indicated by the appearance of feature in the C~{\sc iv} absorption, and
thought to be associated with the interaction between Ly-$\alpha$ photons and
N~{\sc v} ions. We have attempted to explore the reality of these
ghost-features by testing the physically motivated criteria, as set out by
Arav, for the formation of observable features via this mechanism. Of the 21
objects for which we can test all of the criteria, {\em none\/} satisfy all of
the conditions necessary for the formation of an observable feature.  Two of
the criteria, the width of the driving line, and the strength of the UV
continuum are particularly difficult to measure. Of the 5 objects that clearly
pass the latter criteria, having weak He~{\sc ii}, 4 have widths that are
marginally larger than the upper limit imposed by Arav necessary to produce
strong ghost signatures, while the other appears to have only weak
emission-lines.  We recognise that because of the large uncertainty on the
measurements of the He~{\sc ii} strength, 8 further objects from the sample of
21 could potentially satisfy all of the criteria necessary for
ghost-formation.  Addressing this issue will require higher S/N data than
currently available for these objects.

However, taken at face value either the conditions necessary for line-locking
to occur are less stringent than previously proposed, or a large fraction of
these objects are multi-trough interlopers masquerading as ghost-candidates.
This possibility is supported by our finding that the 'fake' ghost zones
bracketing the 'real' ghost zone contain comparable numbers of objects as the
'real' ghost zone. This suggests that the peaks observed within BAL troughs at
the position of the ghost-zone are not a physical effect of radiative line
driving, but are instead independent of any systematic velocity.

Since our analysis has shed real doubt as to the existence of
Ly-$\alpha$-N~{\sc v} line-locking features, in the attached appendix we
repeat for the interested reader, the sequence of tests by Korista
et~al. (1993) and re-investigate the incidence of double-trough phenomena in a
large sample of BALQSOs. In summary, the results of these tests show that
there is no strong evidence for an excess of double troughs bracketing the
region of ghost-formation, and that the evidence for line-locking between
Ly-$\alpha$ photons and N~{\sc v} ions in BALQSO spectra is substantially
lacking.

\onecolumn
%\section{Appendix}
\begin{table}
\begin{tabular}{ l | c | c | c | l | l | c |} \hline \hline
Object Name & Redshift &  Strong N~{\sc v} & Strong Ly~$\alpha$ &
Narrow emission-lines & He~{\sc ii} EW  & FWHM Ghost\\%&Notes
 & & BAL & emission & FWHM (km~s$^{-1}$) & scale factor & (km~s$^{-1})$ \\
\hline 
024514.93-004101.9&2.802&\(\surd\)&\(\surd\)&X(3624)& X 1.06 (0.89--1.31) & 2646\\%&C~{\sc iv} slightly too wide\\
080621.45+191441.0&2.272&\(\surd\)&X&X(4307)& \(\surd\) 1.58 (1.23--2.17) &1785\\%&Strong N V but not C~{\sc iv}\\
083749.59+364145.4&3.416&\(\surd\)&X&\(\surd\)(2807)& \(\surd\) 2.67 (2.05--3.73) &2005\\%&\\
084316.93+050640.1&2.409&\(\surd\)&\(\surd\)&\(\surd\)(2690)& X 0.96 (0.85-1.11) & 1744\\%&\\ 
085104.05+051539.8&3.213&\(\surd\)&\(\surd\)&X(4324)& \(\surd\) 1.56 (1.31--1.93) & 1013\\%&Strangely shaped C~{\sc iv} emission\\
090430.49+343108.9&3.408&\(\surd\)&\(\surd\)&X(4067)& X 0.89 (0.78--1.00) & 1696\\%&\\
092557.52+044035.9&2.266&\(\surd\)&\(\surd\)&X(5219)& \(\surd\) 1.41 (1.09--2.07) & 1248\\
093742.27+403351.2&2.258&\(\surd\)&\(\surd\)&\(\surd\)(2725)& X 1.00 (0.81--1.23) & 1125\\%&\\
094427.27+614424.6&2.337&\(\surd\)&\(\surd\)&\(\surd\)(2426)& X 0.76 (0.88--0.87) & 782\\%&\\
094454.85+481403.5&2.291&\(\surd\)&\(\surd\)&\(\surd\)(2504)& X 0.65 (0.55--0.72) & 1833\\%&\\ 
101844.45+544015.6&3.253&\(\surd\)&\(\surd\)&\(\surd\)(2729)& X 0.43 (0.37--0.53) & 2168\\%&\\
103450.21+313326.1&2.496&\(\surd\)&X&X(3943)& X 1.41 (0.99-2.48) &\\%&\\
105058.28+614406.0&2.797&\(\surd\)&\(\surd\)&\(\surd\)(2375)& X 0.60 (0.53--0.69)  &2185\\%&\\ 
110623.51-004326.0$^{\dagger}$&2.443&\(\surd\)&\(\surd\)&X(4040)& \(\surd\)
3.35 ($>$1.82) & 2996\\%&\\
114704.47+153243.3&3.081&\(\surd\)&\(\surd\)&\(\surd\)(3258)& X 0.89 (0.79--1.00) & 771\\%&\\
131505.19+603716.4&2.330&\(\surd\)&\(\surd\)&X(3976) & X 0.97 (0.83--1.18) & 1313\\%&\\
134818.03+423205.1&3.066&\(\surd\)&\(\surd\)&?& X 0.89 (0.73--1.14) & 944\\%&\\
142244.45+382330.6&3.728&\(\surd\)&\(\surd\)&\(\surd\)(2146)& X 1.16 (0.93--1.51) & 2505\\%&\\
144032.04+060554.1&2.297&\(\surd\)&\(\surd\)&\(\surd\)(2723)& X 0.89 (0.79--1.00) & 1337\\%&\\
162657.47+405848.0&3.051&\(\surd\)&\(\surd\)&\(\surd\)(2520)& X 0.49 (0.44--0.55) & 1333\\%&\\
170322.41+231243.3&2.634&\(\surd\)&\(\surd\)&\(\surd\)(2138)& X 0.93 (0.79--1.05) &1032\\%&\\ 
\hline \\
Object Name & Redshift &  Strong N~{\sc v} & Strong Ly~$\alpha$ &
Narrow emission-lines & He~{\sc ii} EW  & FWHM Ghost\\%&Notes
 & & BAL & emission & FWHM (km~s$^{-1}$) & scale factor & (km~s$^{-1})$ \\%&Notes
\\
005109.45+001636.3&2.036&?&?&X(4850)& X 1.30 (0.94--2.00) &2800\\%&\\
023252.80-001351.1$^{\dagger}$&2.033&?&?&X(4016)& X 0.94 (0.82--1.09) &709\\%&North ghost, Asymmetric emission\\
033048.50-002819.6$^{\dagger}$&1.779&?&?&? & \(\surd\) $>2.00$ &1977\\%&North ghost poor S/N\\
081158.10+343624.1&2.106&?&\(\surd\)&\(\surd\)(3147)& X 1.13 (0.86--1.63) & 4395\\%&\\
082006.60+522158.7&2.015&?&?&\(\surd\)(3217)& ??$^{a}$ & 2660\\%&\\
091109.46+410155.2&2.128&?&\(\surd\)&X(4062)& X 0.56 (0.49--0.64) & 1006\\%&\\
091313.09+411014.2&1.737&?&?&?& \(\surd\) 3.61 (2.62--5.72) & 2607\\%&\\
092221.25+084312.8&1.969&?&?&?& \(\surd\) $>$1.43$^{b}$ & 1398\\%&\\
102908.06+365155.2&1.750&?&?&X(3870)& X 0.31 (0.28--0.34) & 2432\\%&\\
110736.67+000329.4&1.741&?&?&\(\surd\)(2443)& X 1.18 (1.06--1.31) & 2041\\%&\\
111255.70+040600.5&1.989&?&?&X(3706)& X 0.96 (0.85--1.08) & 1055\\%&\\
113831.42+351725.3&2.118&?&\(\surd\)&\(\surd\)(2422)& X 0.78 (0.69--0.89) & 1336\\%&\\  
122107.07+074437.0&1.900&?&?&\(\surd\)(3222)& X 0.91 (0.77--1.08) & 503\\%&\\ 
132304.58-003856.5$^{\dagger}$&1.827&?&?&X(4133)& X 1.09 (0.95--1.28) & 2077\\%&North ghost\\
133428.06-012349.0&1.876&?&?&?& \(\surd\)(??)$^{c}$ & 1616\\%&\\
134458.82+483457.5&2.052&?&?&?& X 0.89 (0.73--1.12) & 1046\\%&\\
141843.95+373750.8&1.782&?&?&X(3596)& X 0.38 (0.33--0.44) &\\%&\\
155505.10+442151.3&1.798&?&?&X(4425)& \(\surd\) 1.90 (1.51--2.59) & 1386\\%&\\
160335.43+225612.9&2.079&?&\(\surd\)&\(\surd\)(2479)& X 0.68 (0.60--0.78) & 2614\\%&\\
163231.60+294929.7&1.905&?&?&\(\surd\)(2587)& X 0.61 (0.54--0.66) & 2140\\%&\\
170056.85+602639.7$^{\dagger}$&2.123&?&\(\surd\)&X(4708)& X 1.15 (0.99--1.35) & 1913\\%&North ghost\\
172001.31+621245.7$^{\dagger}$&1.760&?&?&\(\surd\)(2528)& X 1.73 (1.24--2.95) & 2034\\%&North ghost\\
\hline \\
\end{tabular} 
\caption{Our Ghost sample and the results of testing Arav's criteria along with the FWHM of the ghost feature}
\label{fig:table}
\end{table}
\noindent $^{\dagger}$ Objects identified by North et~al 2006 as their best
ghost-candidates, their Ghost Candidate Final Cut (GCFC).\\
\noindent $^{a}$ Poor continuum fit.\\
\noindent $^{b}$ He~{\sc ii} absorbed.\\
\noindent $^{c}$ He~{\sc II} too weak to be measured.\\

\twocolumn

\onecolumn
\begin{table}
\begin{tabular}{ l | c | c | c | l | l |} \hline \hline
Object Name & Redshift &  Strong N~{\sc v} & Strong Ly~$\alpha$ &
Narrow emission-lines & He~{\sc ii} EW  \\%&Notes
 & & BAL & emission & FWHM (km~s$^{-1}$) & scale factor \\
\hline 
  013724.43-082419.9&2.5663&X&\(\surd\)&\(\surd\)(3174)& X 0.39 (0.35-0.44) \\%\(\surd\)
  014648.52-001051.8&2.3853&\(\surd\)&\(\surd\)&\(\surd\)(2186)& X 0.83 (0.77--1.38 ) \\
  021219.54+141739.1&2.1902&\(\surd\)&X&X(6457)& X 0.94 (0.82--1.18) \\ 
  024413.76-000447.5&2.7925&\(\surd\)&\(\surd\)&X(3804)& \(\surd\)  2.72 ($>$2.46) \\
  031331.22-070422.8&2.7548&\(\surd\)&X&?& \(\surd\) 3.35 (2.43--5.56) \\
  081906.14+394813.8&3.2071&X&\(\surd\)&X(4142)& X 0.43 (0.39--0.47) \\
  084348.68+445226.9&2.5821&X&\(\surd\)&X(4024)& X 1.06 (0.89--1.30) \\
  084554.24+423003.5&2.5568&X&\(\surd\)&\(\surd\)(2378)& \(\surd\) 3.92 ($>$1.69) \\
  090115.18+371822.8&2.6173&X&\(\surd\)&\(\surd\)(2322)& X 0.30 (0.27--0.33) \\
  093804.52+120011.4&2.2273&X&X&?& \(\surd\) 2.03 (1.59--2.79) \\
  095220.88+371622.9&3.0928&\(\surd\)&\(\surd\)&\(\surd\)(3331)& X  0.80 (0.66--1.00) \\
  101324.20+064900.3&2.7675&\(\surd\)&X&?& \(\surd\) $>4.07$ \\
  101420.52+325931.9&2.3935&X&\(\surd\)&\(\surd\)(3256)& X 0.41 (0.37--0.45) \\
  104245.48+365642.2&2.8545&\(\surd\)&\(\surd\)&\(\surd\)(2235)& X 0.44 (0.40--0.50) \\
  110928.51+092403.8&2.1539&X&\(\surd\)&\(\surd\)(3452)& X 0.49 (0.43--0.55) \\
  111437.25+503445.9&2.2075&X&\(\surd\)&X& \(\surd\) $>$3.63 \\
  115901.75+065619.0&2.1906&\(\surd\)&\(\surd\)&X& X 1.02 (0.79--1.43) \\
  135559.03-002413.7&2.3438&X&\(\surd\)&\(\surd\)(3180)& \(\surd\) 1.30 (1.11--1.53) \\
  135912.20+450338.1&2.2794&X&\(\surd\)&X(3802)& X 0.95 (0.68--1.58) \\
  141225.35+041951.9&2.3882&\(\surd\)&\(\surd\)&X(4748)& \(\surd\) 0.72 (0.60--0.89) \\
  143559.60+034153.7&2.4646&\(\surd\)&X&X(4003)& X 1.05 (0.85--1.32) \\
  164148.19+223225.2&2.5061&X&\(\surd\)&\(\surd\)(2347)& X 1.12 (1.00--1.46) \\
  165816.78+231653.7&2.5753&X&\(\surd\)&X(6305)& \(\surd\) 1.88 (1.26--3.66) \\
  214113.05-003545.8&2.2329&X&\(\surd\)&\(\surd\)(3099)& X 0.49 (0.43--0.54) \\
  223841.88+142154.9&2.2898&\(\surd\)&\(\surd\)&\(\surd\)(2488)& X 0.47 (0.43--0.52) \\
\hline \\
\end{tabular}
\caption{As for Table~1, indicating our comparison sample of 25 objects
  selected to have broad absorption and no ghost-features and the results of
  testing Arav's criteria for ghost formation.}
\label{fig:Comptable}
\end{table}
\twocolumn

\onecolumn

\begin{figure}
\centering \includegraphics[width=0.95\textwidth]{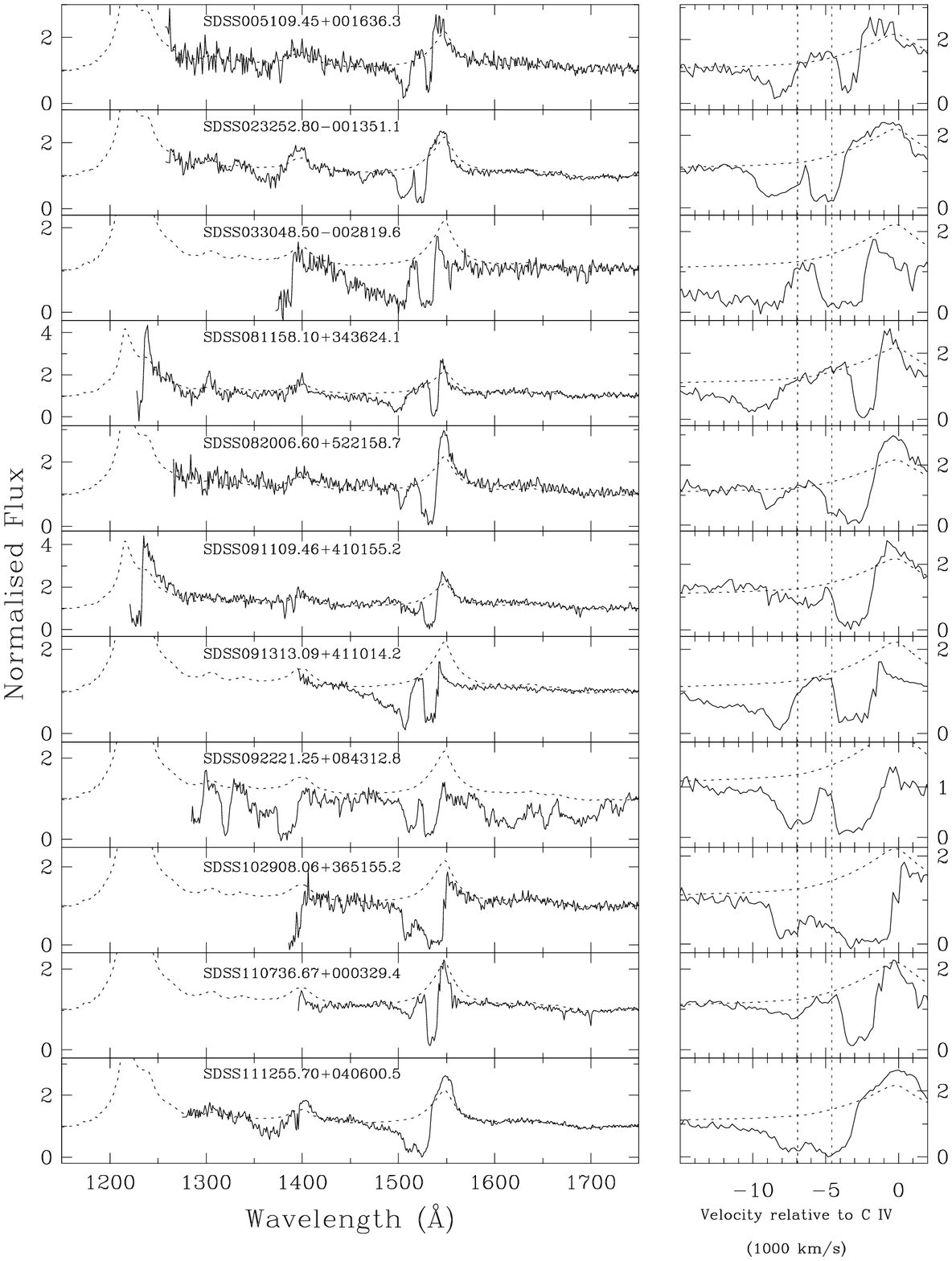}
%height=0.95\textheight]{NOTFWHMa}
\caption{Ghost candidates for which only some of the criteria can be tested.}
\label{fig:mike_3b}
\end{figure}
\begin{figure}

\centering \includegraphics[width=0.95\textwidth]{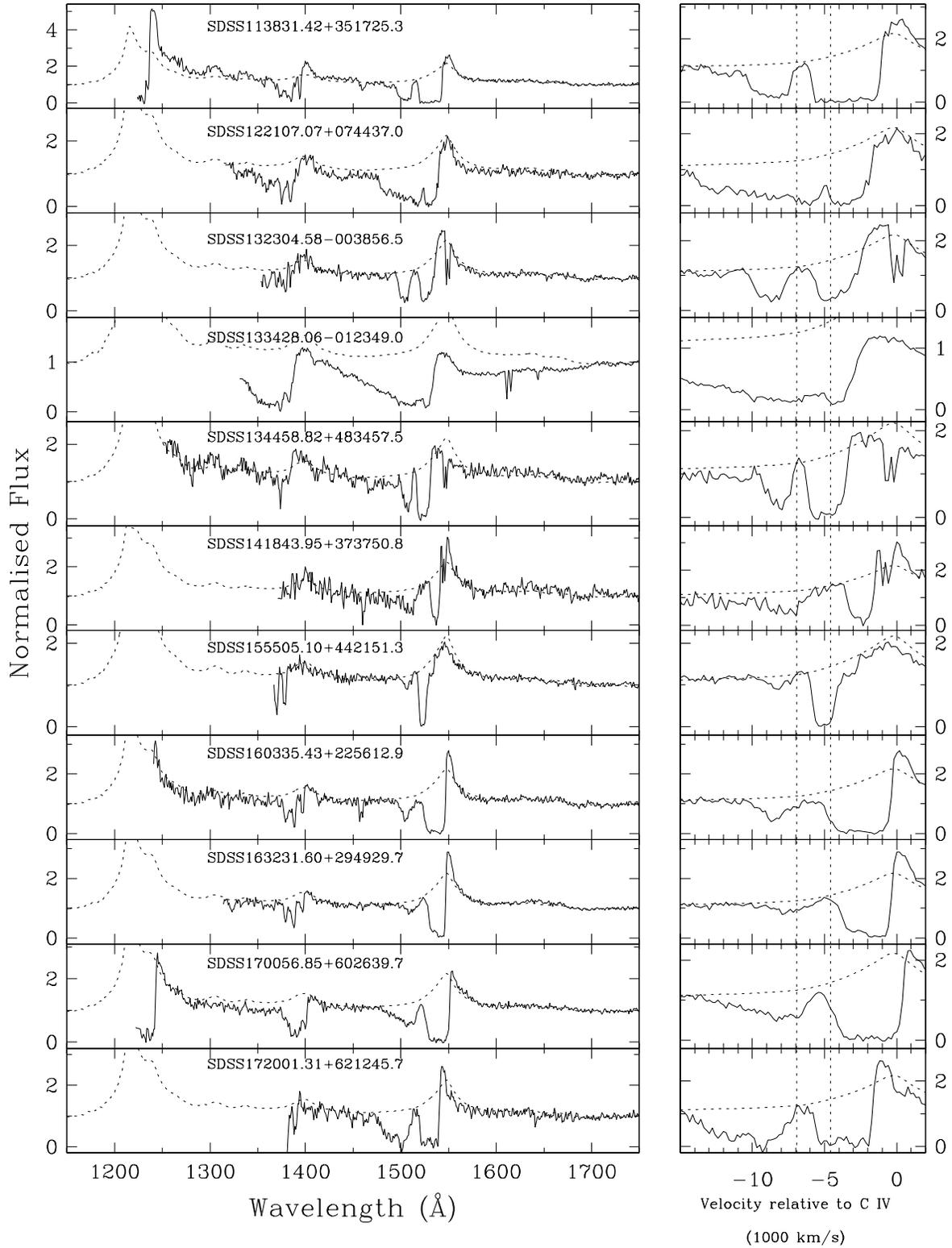}
%height=0.95\textheight]{NOTFWHMa}
\caption{Ghost candidates for which only some of the criteria can be tested.}
\label{fig:mike_4b}
\end{figure}
\twocolumn

\section*{Acknowledgements}

\label{lastpage}

\appendix
\section{On the reality of the double-trough phenomena}

Since the criteria by which observable ghosts are formed appears less
stringent than has previously been claimed, we have performed a series of
sanity checks to test whether ghost-features and double-troughs in general,
are not simply a result of the random superposition of multiple absorption
systems. The tests performed are similar to those described in Korista
et~al. (1993), though we note that our parent BALQSO sample is considerably
larger and thus the quality of the statistics is a significant improvement
over previous work.

\subsection{Random sampling}

In the first test we compute the geometric mean residual intensity spectrum
(the ratio of the geometric mean BALQSO composite divided by a geometric mean
non-BALQSO composite spectrum) from two random samples (without replacement)
taken from the BALQSO catalogue of Scaringi et~al. (2009) (Figure~A1, sample 1
(solid line), sample 2 (dashed line).  These samples (consisting of 1776
objects each) were generated by sorting the BALQSO catalogue into RA order,
and then selecting even-numbered objects for the first sample, and
odd-numbered objects for the second sample.  The mean residual intensity
spectra for both samples are virtually indistinguishable, aside from a small
difference ($\sim$15\%) in the residual intensity in the velocity range
3900--7800~km~s$^{-1}$ blue-ward of line centre, which intriguingly, is
precisely where the purported ghost feature should be found.  We have checked
to see whether there exists an excess of ghost-candidate spectra in either
sample, and find that the ghost candidates are spread evenly amongst both
samples, with 21 in sample A and 22 in sample B.  While there are small
apparent differences in the mean residual intensity spectrum in the region of
the ghost-feature, neither sample shows convincing evidence of a double-trough
feature.

\begin{figure}
\centering \includegraphics[width=0.5\textwidth]{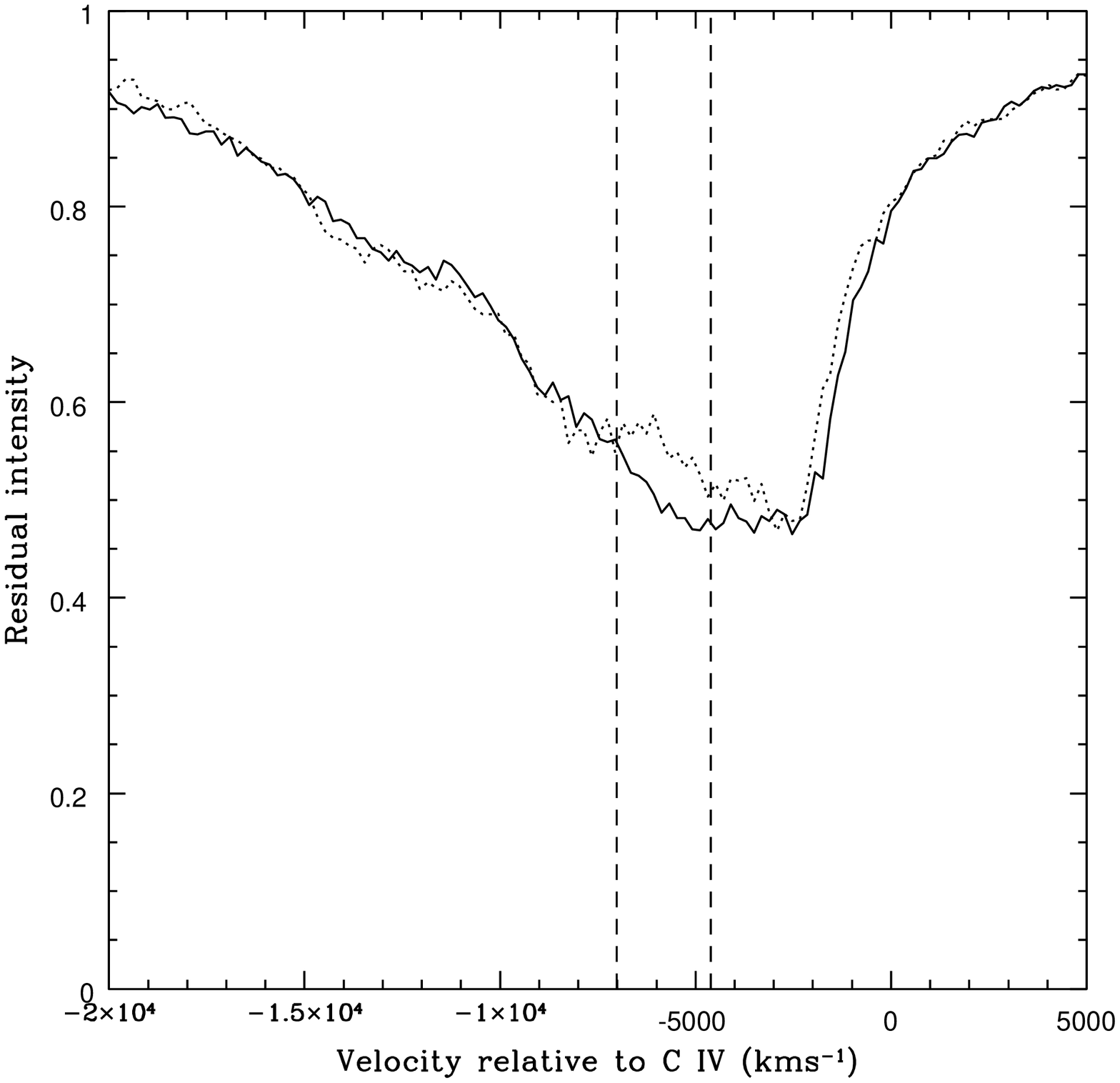}
\caption{Geometric mean composite residual intensity spectrum of two randomly 
selected  BALQSO samples (without replacement).}
\label{fig:090630_mt_2}
\end{figure}

To explore these findings further, we have attempted to address whether there
exists any substantive evidence for absorption components at preferred
velocities. If ghost features are relatively common, we would expect an excess
abundance of peaks within the absorption trough at velocities 5900~km/s
blue-ward of line centre.  To address this question we have performed two
additional tests. In these tests we work with fluxes instead of residual
intensities to avoid complications introduced by uncertainties in fitting the
C~{\sc iv} emission-line blue-ward of line centre.  In the first, we take the
LVQ BALQSO sample of Scaringi et~al. 2009, and compare the fluxes in 3
adjacent velocity bins (A, B and C) each with widths of 1000~km/s, where B is
the reference bin and A and C represent the adjacent bins. Sliding these bins
from a velocity of 0--30,000 km/s blue-ward of C{\sc iv } rest, we record the
number of times the mean flux in the central bin (bin B) exceeds the mean flux
in adjacent bins A and C by more than 3 $\sigma$. To show the effects of
selecting different bin-widths over which the fluxes are measured, we repeat
this exercise for bin widths of 500 and 1500 km/s.  The results of this
procedure are shown in Fig~A2 (upper left panel). The distribution of peaks
shows a prominent peak at zero velocity, which we associate with the location
of the C~{\sc iv} emission-line peak. Ignoring this feature, the peak
distribution is a generally smooth function of velocity with no indication of
an excess number of peaks at 5900 km/s blue-ward of line centre.  The same
analysis performed on our ghost candidate sample, displays a similar peak at
zero velocity, and in addition further peaks at around 6000 km/s and
11,000~km/s. While the numbers of peaks at higher velocities is only
marginally less than that at 5900 km/s, visual inspection of our ghost
candidate sample suggest that the significance of these secondary peaks is
rather low.

We have also performed a variation on this method in which instead of
measuring the intensities within each bin, we measure the mean gradient of the
intensities within each of the bins. A peak is recorded if and only if the
gradient in bin A is positive, the gradient in bin A is greater than the
gradient in bin B, the gradient in bin B is greater than gradient of C, and
the gradient in bin C is negative. Figure~A2 lower panels shows the result of
this test.  Again, the peak at zero velocity we associate with the peak of the
C~{\sc iv} emission-line.  Thereafter the distribution of the locations of the
peaks follows a smooth broad function with a maximum at approximately 15,000
km/s.  As with the previous test, varying the bin widths between 500-1500 km/s
does not significantly alter the results.

These additional tests suggest that for the general BALQSO population, there
is no substantive evidence that there exist preferred velocities for the
location of peaks within the broad absorption troughs of BALQSOs. Thus if
ghosts are real they are certainly rare, as was first suggested by Arav.  We
note that the incidence of single peaks (or double troughs) in our sample
258/3552 (7\%) is significantly smaller than that found by Korista
et~al. (1993, 22\%), likely a result of the small number statistics of this
earlier study.

\onecolumn
\begin{figure}
\centering \includegraphics[width=0.7\textwidth,angle=270]{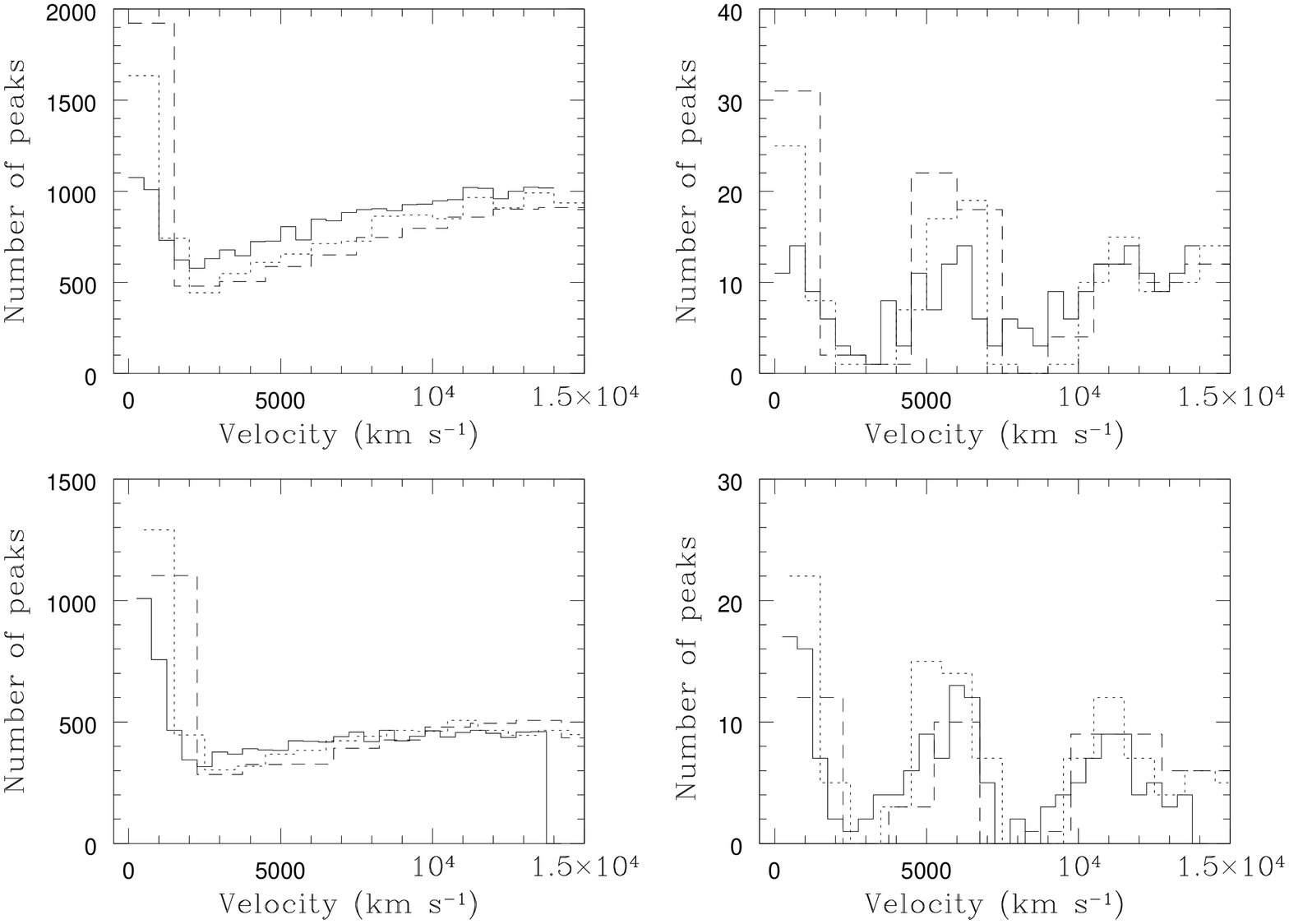}
\caption{Distribution functions showing the incidence of peaks as a function
of velocity blue-ward of C{\sc iv} rest wavelength (see text for
details). Upper left panel - LVQ BALQSOs, peak distribution functions (solid
line 500 km/s bins, dotted line 1000 km/s bins, dashed lines 1500 km/s
bins. Upper right panel - as for upper left for the ghost-candidate sample.
Lower panels , as above except now  using slope comparisons.}
\label{fig:plot_peaks}
\end{figure}

\twocolumn

\subsection{Appearance of double-troughs at other velocities}

In the second test we verify whether there exist associated troughs (via
line-locking) at velocities other than systemic. We do this by aligning each
spectrum according to the location of the red-edge of the first BAL trough and
thereby search for evidence of a bi-modal distribution of sub-troughs. Here,
the red-edge is defined as in Korista et~al. (1993) to be where the residual
intensity falls below 0.75 and remains below this level for more than
1000~km~s$^{-1}$. The resultant geometric mean residual intensity spectrum is
shown in Figure~A3. As one might expect, the residual intensity is somewhat
steeper at lower velocities, a result of aligning the first troughs in
velocity space. However, the remainder of the spectrum is a smooth function of
velocity, with no indication of additional sub-troughs. Thus, as with Korista
et~al. (1993) we find no evidence for line-locking features at velocities
other than systemic.

% aim to is to check for a bimodal distribution of sub-troughs
% claim is that line-locking is not linked to the velocity of the first trough

% ++++ distribution of velocities of first trough show no preferred velocities

Unlike Korista et~al. (1993) the distribution of velocities of the first
trough peaks at velocities between 1500--2000~km~s$^{-1}$ blue-ward of
line-centre, thereafter the number of objects contributing to each velocity
bin shows a smooth linear decline. Thus while there appears a preference for
the onset of BALs toward lower velocities, there is no indication of other
preferred blue-ward velocities.  This again suggests that the incidence of the
double-trough signature among the BALQSO population as a whole is generally
low.  We note that the distribution of redshifted troughs are similarly peaked
toward lower velocities and may indicate a preferred launch radius for the
out-flowing gas.

\begin{figure}
\centering \includegraphics[width=0.5\textwidth]{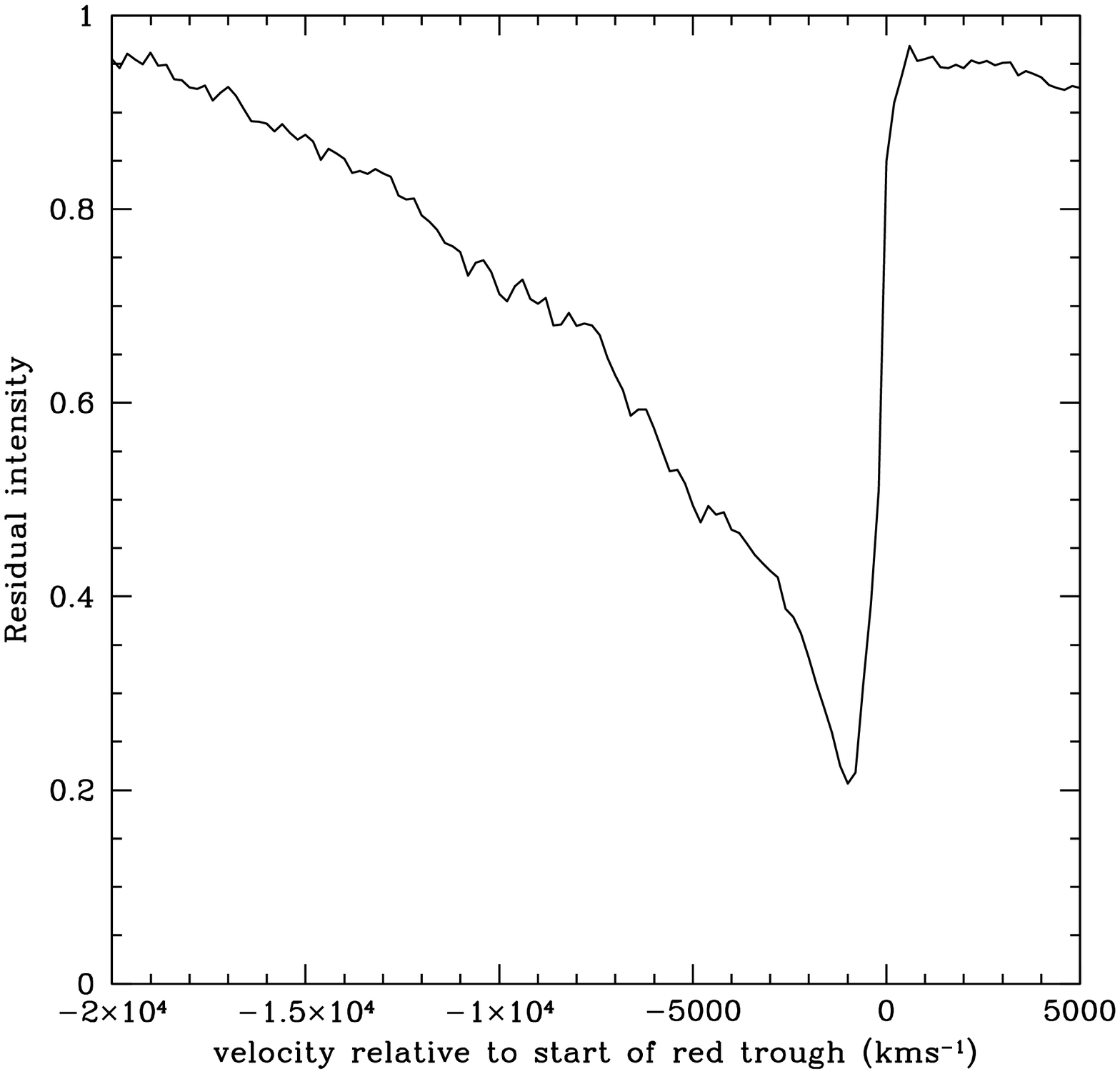}
\caption{Residual intensity spectra after aligning all spectra to the position
  of the first (lowest velocity) absorption trough. We see no indication of a
  second trough at the known separation of Ly-$\alpha$--N~{\sc v}.}
\label{fig:090630_mt_red}
\end{figure}

\begin{figure}
%\centering \includegraphics[width=0.5\textwidth]{090703_vel_trough}
\centering \includegraphics[width=0.5\textwidth]{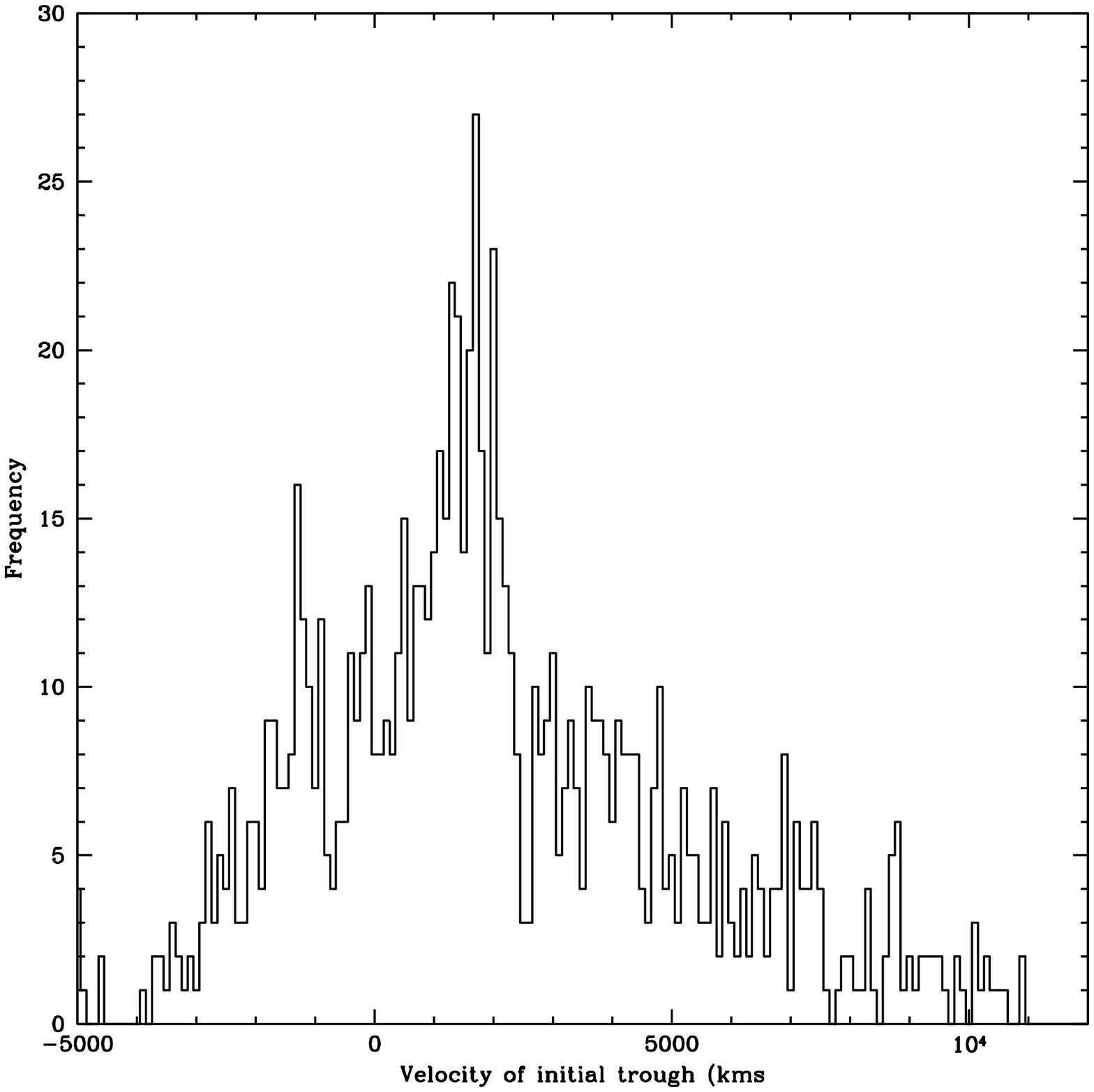}
\caption{The frequency distribution of velocities of the first absorption
trough.}
\label{fig:090703_vel_trough}
\end{figure}

\begin{figure}
\centering \includegraphics[width=0.5\textwidth]{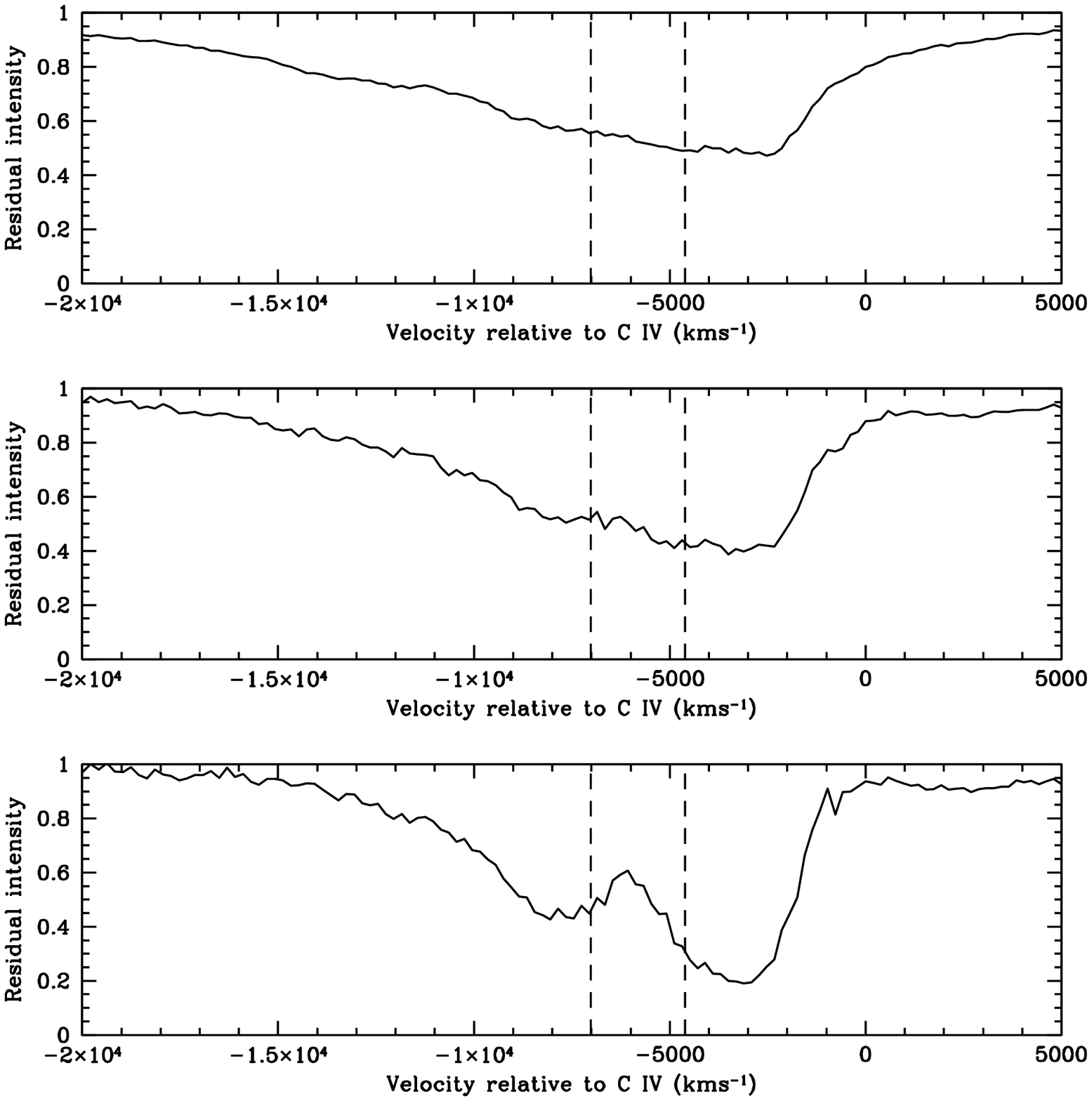}
\caption{Residual intensity spectra of (i) the multi-trough sample, (ii)
the single peak sample, and (iii) ghost-candidate sample. See text for details}
\label{fig:090630_res_comps}
\end{figure}

\subsection{K-S test}
In our third test, we compute the mean residual intensities in 3 separate
velocity bins, one centred on the expected location of any ghost-feature, the
other taken from bins either side of this feature, ie. velocity bins at
4000--5000~km~s$^{-1}$, 5500--6500~km~s$^{-1}$ (the reference bin), and
10000--11000~km~s$^{-1}$ respectively. We then use the Kolmogorov-Smirnov (KS)
test to check whether the distribution of residual intensities within each of
these velocity bins belongs to the same population as the reference bin. The
residual intensity spectra we use for this test are (i) the total sample of
BALQSOs from Scaringi et~al. (2009), the randomly selected samples 1 and 2
described above, our Multi-trough sample from rejection cut 2, our single peak
sample from rejection cut 3 and our candidate ghost sample. The results of
this test are presented in Table~A1.

Table~A1 gives the probability that the K-S statistic of the sub-trough with
the reference trough (i.e. the ghost zone) would be exceeded by chance if they
were drawn from the same population. That is, lower K-S probabilities indicate
more significant differences between the chosen velocity bin and the reference
bin.  Our analysis shows that only for our ghost sample do {\em both\/}
sub-troughs 1 and 2 indicate significant differences from the reference bin
(K-S probabilities less than 0.1\%). This is as expected, since members of
this sample were selected precisely because they showed significant ghost-like
features. The single peak sample shows no evidence that the residual intensity
of the ghost zone is drawn from a different population to that of sub-trough
2. This suggests that the single peak sample does not have a systematic peak
and that the position of peaks within the absorption is random and unrelated
to Ly-$\alpha$-N~{\sc v} line locking.

\begin{center}
\begin{table}
\begin{tabular}{lccc} \hline \hline
& & \multicolumn{2}{c}{K-S probabilities $\times$ 100\%} \\
BALQSO Sample & Sample  & Sub-trough  & Sub-trough \\
 Sample & Size &  1 & 2 \\
\hline
LVQ BALs                 & 3552 & 10$^{-2}$ & 5.9 \\
MTS   & 1019 & 5$\times$10$^{-3}$ & 23.8 \\
MTS A &  510 & 1.2 & 34.9 \\
MTS B &  509 & 2.1 & 4.1 \\
Single Peak Sample       &  258 & 3.7 & 41.4 \\
Ghost Zone Sample        &   69 & 0.2 & 0.1 \\
\hline 
\end{tabular}
\caption{K-S tests on the multi-trough phenomenon}
\end{table}\label{kstest}
\end{center}

\subsection{A Monte Carlo test for ghosts}

Korista et~al. (1993) described one further test for an excess of
double-trough features bracketing the Ly-$\alpha$-N~{\sc v} line-locking
region, which we repeat here for completeness.  In this Monte Carlo
simulation, we test how frequently the depth of the double-trough structure is
exceeded by a random arrangement of residual intensity differences among each
of the 3 velocity bins (see Korista et~al. 1993 for details). For each
spectrum, we can form 6 residual intensity differences from 3 velocity
bins. For each spectrum in our BALQSO catalogue we randomly pick 2 from the 6
possible residual intensity differences and calculate their mean value,
i.e. an average residual intensity difference. We then define $<D>_{random}$
to represent the mean of these average intensity differences across our BALQSO
sample. $<D>_{random}$ is then compared to $<D>$, formed by averaging the mean
residual intensity differences between sub-troughs 1 and 2 with the reference
bin, again averaged across our whole sample.  $<D>$ can take on any value
between $\pm1$. Values of $<D>$ near $+1$ indicate a strong double-trough
structure. Conversely, a value of $<D>$ near $-1$ indicates a strong trough at
the position of the reference bin. Values near zero are indicative of a smooth
trough passing through the velocity bins.

To test how frequently a double-trough structure having a contrast at least as
large as that observed in the sample mean could arise by chance, we repeat
this process 10,000 times for each sample, and determine the number of times
$<D>_{random}$ exceeds $<D>$, normalised to 10,000. The results of these
simulations are shown in Table~A2.
\begin{center}
\begin{table}
\begin{tabular}{lcc} \hline \hline
BALQSO Sample & $<D>$ & Monte Carlo \\
 & & Probability x 100\% \\
\hline
LVQ BALs&-0.01197&98.88\\
MTS &-0.03918&99.8\\
Single Peak Sample&0.007355&35.8\\
Ghost Zone Sample&0.16932&0.0\\
\hline
\end{tabular}
\caption{Monte Carlo simulations of the double trough phenomenon}
\end{table}\label{montecarlo}
\end{center}
\noindent Since we only find strong evidence for the double-trough structure
within our ghost zone sample (chosen for precisely that reason), we conclude
that there is indeed no strong evidence for an excess of objects with
sub-troughs bracketing the ghost-zone among the BALQSO population at large.
Indeed, the high probabilities found for both the LVQ sample and
for the multi-trough sample, strongly suggest that multi-trough features are 
randomly distributed in velocity. We are therefore conclude that 
many of the objects previously identified as strong ghost-candidates
in previous studies may simply be MT interlopers masquerading as ghosts due 
to the chance alignment of multiple un-associated absorption systems.

\subsection{SDSS J101056.68+355833.3: Si IV ghost?}

\begin{figure}
\centering \includegraphics[width=0.5\textwidth]{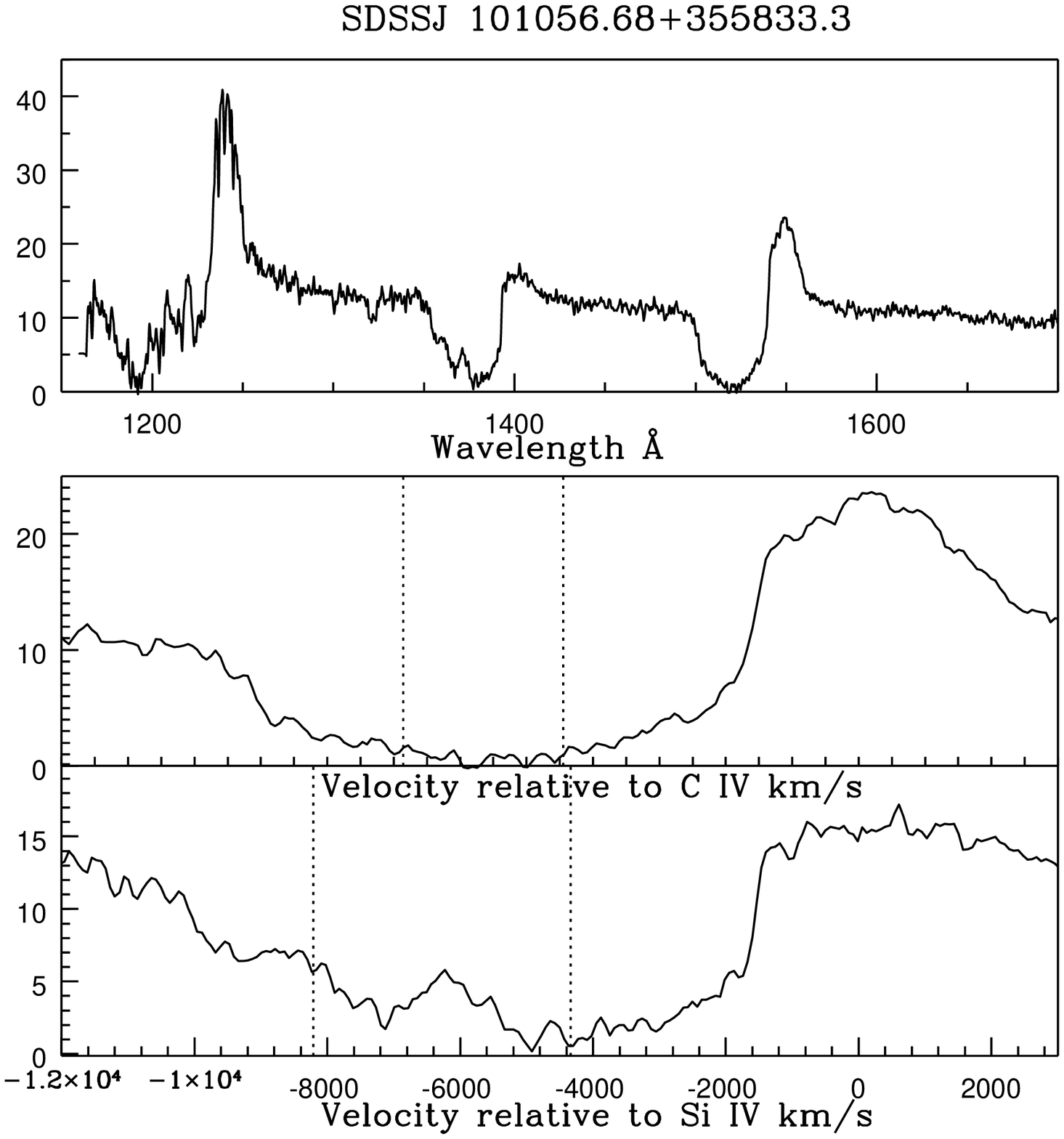}
% height=0.3\textheight]{SiIVghost}
\caption{SDSS J101056.68+355833.3: A possible Si~{\sc iv} ghost showing no
evidence of a C~{\sc iv} feature along with close ups in velocity space of the
C~{\sc iv} (middle) and Si~{\sc iv}(top) ghost regions. Vertical lines
represent the ghost zone}
\label{fig:SiIV}
\end{figure}

This is an object selected for our comparison sample as it shows no hint of a
ghost feature within the deep broad C~{\sc iv} absorption. The spectra is
shown in Figure~\ref{fig:SiIV} along with the flux in velocity space relative
to C~{\sc iv} and Si~{\sc iv} showing clearly the potential ghost feature in
the Si~{\sc iv} absorption. SDSS~J101056.68+355833.3 meets the criteria for
strong emission due to C~{\sc iv} and N~{\sc v} emission exceeding 100\% of
the continuum flux at their peak, shows clear N~{\sc v} absorption, has C~{\sc
iv} FWHM measured as 3483~km~s$^{-1}$ and an EW of He~{\sc ii} measured to be
3.58$\pm$0.82~\AA. These measurements place this object very close to the
boundary on several criteria and it would not be surprising to see a ghost
feature in this object. While no ghost feature can be seen in the C~{\sc iv}
trough, there is a clear feature within the Si~{\sc iv} trough
($\sim$6200km~s$^{-1}$) within the ghost zone. Due to the greater separation
of the Si~{\sc iv} doublet(1933~km~s$^{-1}$) in comparison to the C~{\sc iv}
doublet(498km~s$^{-1}$) a ghost feature in the Si~{\sc iv} would be expected
to be wider and weaker. This larger doublet separation is the reason for the
wider ghost zone in the S~{\sc iv} BAL trough. The absence of a C~{\sc iv}
feature could be due to an extremely high optical depth in the flow such that
the flow remains optically thick even with the decrease in optical depth
caused by the acceleration due to N~{\sc v} ions. However the absorption
troughs of C~{\sc iv} and Si~{\sc iv} look very similar with the exception of
the potential ghost feature suggesting the optical depths are similar. There
is a slight hint of an additional feature within the Si~{\sc iv} at
$\sim$8200km~s$^{-1}$ which is in the appropriate region to be due to the
lower wavelength doublet feature from Si~{\sc iv}.

%\onecolumn
%\twocolumn

\end{document}